\PassOptionsToPackage{dvipsnames,table}{xcolor}
\documentclass[sigconf,screen,pbalance,hidelinks,natbib=true]{acmart}

\usepackage{microtype} %

\usepackage{url}
\usepackage{hyperref}

\usepackage{xspace}

\usepackage{graphicx}
\usepackage{caption}
\usepackage{subcaption}
\usepackage{multirow}
\usepackage{comment}
\usepackage{adjustbox}
\usepackage{enumitem}

\newcommand\ag[1]{\textcolor{black}{#1}}

\usepackage[textsize=tiny]{todonotes}

\newcommand{\method}[0]{EarlyBIRD}
\newcommand{\CLS}{\texttt{CLS}\xspace}

\newcommand{\head}[1]{\par\noindent\textbf{#1:}\space}
\newcommand{\subhead}[1]{\par\noindent\emph{#1}\space}

\usepackage{fancybox}
\usepackage[framemethod=tikz]{mdframed}
\mdfdefinestyle{mystyle}{%
  rightline=false,%
  innerleftmargin=5,
  innerrightmargin=5,
  outerlinewidth=1pt,
  topline=false,
  leftline=false,
  bottomline=false,
  skipabove=\topsep,
  skipbelow=\topsep,
  backgroundcolor=RoyalBlue!10
}

\AtBeginDocument{%
  \providecommand\BibTeX{{%
    \normalfont B\kern-0.5em{\scshape i\kern-0.25em b}\kern-0.8em\TeX}}}

\received{2023-02-02}
\received[accepted]{2023-07-27}
\makeatletter
\def\@copyrightpermission{
  \hspace*{0mm}\includegraphics[width=2cm]{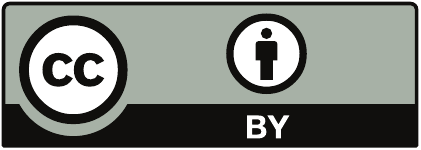}%
  \hspace*{2mm}\raisebox{2.5mm}[25pt][5pt]{%
          \parbox{\columnwidth}{\footnotesize This work is licensed under a Creative Commons \\ Attribution 4.0 International (CC BY 4.0) license.}%
  }\newline%
  The content in this pre-print is the same as in the CRC accepted for publication in:
}%
\makeatother
\acmDOI{10.1145/3611643.3616304}
\setcopyright{rightsretained}
\acmPrice{}
\acmISBN{}
\acmYear{2023}
\copyrightyear{2023}
\acmConference[ESEC/FSE '23]{Proceedings of the 31st ACM Joint European Software Engineering Conference and Symposium on the Foundations of Software Engineering}{December 3--9, 2023}{San Francisco, CA, USA}

\begin{document}

\title[The EarlyBIRD Catches the Bug: On Exploiting Early Layers of Encoder Models\ldots]{The EarlyBIRD Catches the Bug: On Exploiting Early Layers of Encoder Models for More Efficient Code Classification}

\author{Anastasiia Grishina}
\email{anastasiia@simula.no}
\orcid{0000-0003-3139-0200}
\affiliation{
  \institution{Simula Research Laboratory}
  \city{Oslo}
  \country{Norway}
}

\author{Max Hort}
\email{maxh@simula.no}
\orcid{0000-0001-8684-5909}
\affiliation{
  \institution{Simula Research Laboratory}
  \city{Oslo}
  \country{Norway}
}

\author{Leon Moonen}
\email{leon.moonen@computer.org} %
\orcid{0000-0002-1761-6771}
\affiliation{
  \institution{Simula Research Laboratory \& \newline 
               BI Norwegian Business School}
  \city{Oslo}
  \country{Norway\\[1.5ex]}
}

\newcommand\ana[1]{\textcolor{purple}{#1}}
\newcommand\leon[1]{\textcolor{PineGreen}{#1}}
\newcommand\maxh[1]{\textcolor{blue}{#1}}
\renewcommand\ana[1]{\textcolor{black}{#1}}
\renewcommand\leon[1]{\textcolor{black}{#1}}
\renewcommand\maxh[1]{\textcolor{black}{#1}}

\begin{abstract}
The use of modern Natural Language Processing (NLP) techniques has shown to be beneficial for software engineering tasks, 
such as vulnerability detection and type inference.
However, training deep NLP models requires significant computational resources.
This paper explores techniques that aim at achieving the best usage of resources and available information in these models.

We propose a generic approach, \method, to build composite representations of code from the early layers of a pre-trained transformer model. 
We empirically investigate the viability of this approach on the CodeBERT model by comparing the performance of 12 strategies for creating composite representations with the standard practice of only using the last encoder layer. 

Our evaluation on four datasets shows that several early layer combinations yield better performance on defect detection, and some combinations improve multi-class classification. 
More specifically, we obtain a +2 average improvement of detection accuracy on Devign with only 3 out of 12 layers of CodeBERT and a 3.3x speed-up of 
fine-tuning.
These findings show that early layers can be used to obtain better results using the same resources, 
as well as to reduce resource usage during fine-tuning and inference.

\end{abstract}



\maketitle

\section{Introduction}

Automation of software engineering (SE) tasks supports developers in 
creation and maintenance of source code. 
Recently, deep learning (DL) models have been trained on large open-source code corpora and used to perform code analysis tasks~\cite{chen2021:evaluating, allamanis2018:survey, sharma2021:survey, lu2021:codexglue}.
Motivated by the naturalness hypothesis stating that code and natural language share statistical similarities, 
researchers and tool vendors have started training deep NLP models on code and fine-tuning them on SE tasks~\cite{devanbu2015:new}.
Amongst others, such models have been applied to type inference~\cite{hellendoorn2018:deep}, code clone detection~\cite{zhao2018:deepsim}, 
program repair~\cite{ye2022:neural,chen2019:sequencer,yasunaga2021:breakitfixit,fu2022:vulrepair}, 
and defect prediction~\cite{russell2018:automated,chakraborty2021:deep,wei2021:contextaware,pan2021:empirical}.
In NLP-based approaches, SE tasks are frequently translated to code classification problems. 
For example, 
detection of software vulnerabilities is a binary classification problem, bug type inference is a multi-class classification setting, and type inference is a multi-label multi-class classification task in case a type is predicted for each variable in the program. 

Most modern NLP models build on the transformer architecture~\cite{vaswani2017:attention}. 
This architecture uses \emph{attention} mechanism and consists of an \emph{encoder} that converts an input sequence to a \emph{representation} through a series of layers, 
followed by \emph{decoder} layers that convert this representation to an output sequence. 
Although effective in terms of learning capabilities, 
the transformer design results in multi-layer models that need large amounts of data for training from scratch. 
A well-known disadvantage of these models is the high resource usage that is required for training due to both model and data sizes.
While a number of pre-trained models have been published recently, 
fine-tuning these models for specific tasks still requires additional computational resources~\cite{lu2021:codexglue}. 

This paper explores techniques that aim at optimizing the use of resources and information available in models during fine-tuning. 
In particular, we consider open white-box models, for which the weights from each layer can be extracted.
We focus on encoder-only models, as they are commonly used for SE classification tasks, in particular, the transformer-based encoders. 
The standard practice in encoder models is to obtain the representation of the input sequence from the last layer of the model~\cite{feng2020:codebert}, 
while information from earlier layers is usually discarded~\cite{karmakar2021:what}. 
\leon{I.e., while the early layers are used to \emph{compute} the values of the last layer, they are generally not considered as individual representations of the input in the way that the last layer is.}
To exemplify the amount of discarded information at inference,  
when fine-tuning a 12-layered encoder, such as CodeBERT~\cite{feng2020:codebert}, for bug detection, 
92\% of the code embeddings are ignored.\footnote{~\leon{That is, the weights from 11 out of 12 layers are ignored for classification.}}
However, it has been shown for natural language that the early layers of an encoder capture lower-level 
syntactical features better than the later layers~\cite{blevins2018:deep, peters2018:dissecting, liu2019:linguistic, sun2020:how:arxiv}, 
which can benefit downstream tasks.

Inspired by the line of research that exploits early layers of models, 
we propose \method{},\footnote{~Early-layer Based Improvement or Reduction of resources useD} 
a novel and generic approach for building composite representations %
from the early layers of a pre-trained encoder model.
\method{} aims to leverage all available information in existing pre-trained encoder models during fine-tuning to either improve results or achieve competitive results at reduced resource usage during code classification. 
We empirically evaluate \method{} on CodeBERT~\cite{feng2020:codebert}, a popular pre-trained encoder model for code, 
and four benchmark datasets \ana{that cover three \leon{common} SE tasks:} defect detection with the Devign and ReVeal datasets~\cite{zhou2019:devign, kanade2020:learning}, 
bug type inference with the data from Yasunaga et al.~\cite{yasunaga2021:breakitfixit}, and exception type classification~\cite{chakraborty2021:deep}.
The evaluation compares the \emph{baseline} representation that uses the last encoder layer with results obtained via \method{}. 
We both fine-tune the full-size encoder and its pruned version with only several early layers present in the model.
The latter scenario analyzes the trade-off between only using a partial model and the performance impact on SE tasks.

\head{Contributions} In this paper, we make the following contributions:

\subhead{(1) We propose \method{}, an approach for creating composite representations of code using the early layers of a transformer-based encoder model.} 
The goal is to achieve better code classification performance at equal resource usage or comparable performance at lower resource usage.
    
\subhead{(2) We conduct a thorough empirical evaluation of the proposed approach.} 
We show the effect of using composite \method{} representations while fine-tuning the original-size CodeBERT model on four real-world code classification datasets.
We run \method{} with 10 different random initializations of non-fixed trainable parameters and mark the \method{} representations that yield statistically significant improvement over the baseline. 
    
\subhead{(3) We investigate resource usage and performance of pruned models.} 
We analyze the trade-off between removing the later layers of a model and the impact this has on classification performance.

\head{Main findings}
With \method{}, we achieve performance improvements over the baseline code representation with the majority of representations obtained from single early layers on the defect detection task and selected combinations on bug type and exception type classification. 
Moreover, out of the reduced-size models with pruned later layers, we obtain a +2 average accuracy improvement on Devign with 3.3x speed-up of fine-tuning, as well as +0.4 accuracy
improvement with 3.7x speed-up on average for ReVeal.

The remainder of the paper is organized as follows. We present related work in Section~\ref{sec:related-work} and provide background details of the study in Section~\ref{sec:background}. The methodology is described in Section~\ref{sec:methodology} which is followed by experimental setup in Section~\ref{sec:experiment}. 
We present and discuss results in Section~\ref{sec:results} and conclude with Section~\ref{sec:conclusion}.

\section{Related Work}
\label{sec:related-work}
Here, we give an overview of language models for SE tasks and recent encoder models, specifically, as well as different approaches to use early layers of encoder models. 

\subsection{Transformers in Software Engineering}
The availability of open source code and increased hardware capabilities popularized training and usage of 
\ana{Deep Learning, including NLP and Large Language Models (LLMs), for}
SE tasks. 
To date, deep NLP models have already been 
applied in at least 18 SE tasks~\cite{niu2022:deep}.  
Pre-trained language models available for fine-tuning on SE tasks largely build on the transformer architecture, sequence-to-sequence models, and the attention mechanism~\cite{vaswani2017:attention,chen2019:sequencer,chen2021:evaluating}. 
One widely used benchmark to test different deep learning architectures on SE tasks is CodeXGLUE~\cite{lu2021:codexglue}. 
The benchmark provides data, source code for model evaluation, and a leader-board ranking model performance on different tasks~\cite{lu2021:codexglue}.

SE tasks can be translated to input sequence classification and generation of code or text. 
Examples of generative tasks in SE are code completion, code repair, generation of documentation from code and vice versa, and translation between different programming languages. 
Such tasks are frequently approached with neural machine translation models. 
Full transformer models for translation from a programming language (PL) to a natural language (NL) or PL-PL tasks include PLBART~\cite{ahmad2021:unified}, PYMT5~\cite{clement2020:pymt5}, TFix~\cite{berabi2021:tfix}, CodeT5~\cite{wang2021:codet5}, Break-It-Fix-It~\cite{yasunaga2021:breakitfixit}. %
Alternatively, generative models can include the decoder-only part of the transformer as in GPT-type models. 
In this case, the decoder both represents the input sequence and transforms it into the output sequence. 
Decoder-based models for code include, for example, Codex and CodeGPT~\cite{chen2021:evaluating,lu2021:codexglue}.

In the tasks that require code or documentation representation and their subsequent classification, the encoder-only architectures are used more frequently than in translation tasks.
Examples of code classification problems are code clone detection, detection of general bugs, such as the presence of swapped operands, wrong variable names, syntax errors, or security vulnerabilities. 
A number of encoder models for code applied a widely-used bi-directional encoder, BERT~\cite{devlin2019:bert}, to pre-train it on code, with some modifications of the input.  
In this way, the CodeBERT~\cite{feng2020:codebert}, GraphCodeBERT~\cite{guo2021:graphcodebert}, CuBERT~\cite{kanade2020:learning}, and PolyglotCodeBERT~\cite{ahmed2022:multilingual} models were created. 
In detail, the 12-layer RoBERTa-based CodeBERT model was pre-trained on NL-PL tasks in multiple PLs and utilized only the textual features of code. 
Note that RoBERTa is a type of BERT model with optimized hyper-parameters and pre-training procedures~\cite{liu2019:roberta}. 
Together with the decoder-only CodeGPT model, the encoder-only CodeBERT model was used as a baseline in CodeXGLUE. 
GraphCodeBERT utilizes both textual and structural properties of code to encode its representations. 
PolyglotCodeBERT is the approach that improves fine-tuning of the CodeBERT model on a multi-lingual dataset for a target task even if the target task tests only one PL. 
This paper focuses on the fine-tuning strategies which, in contrast to PolyglotCodeBERT, do not increase the resource usage for fine-tuning.
CuBERT is a 24-layer pre-trained transformer-based encoder tested on a number of code classification tasks, including exception type classification.
We test the performance of the proposed \method{} composite representations on defect detection, including the use of one of CodeXGLUE benchmarks, as well as on error and exception type classification tasks.
However, the goal of this paper is to achieve improvement over the baseline model when it is fine-tuned with composite code representations. We do not aim to compare results with other models, but rather propose an approach that is applicable to transformer-based encoders for source code and show its performance gains compared to the same model usage without the proposed approach.

\subsection{Use of Early Encoder Layers}

A number of studies explored different approaches to use information from early layers of DL models for sequence representation, such as probing single layers, pruning and variable learning rates. 
One way to leverage information from early model layers is to give different priority to layers while fine-tuning the models ~\cite{howard2018:universal,sun2019:how}.
For example, the layer-wise learning rate decay (LLRD) strategy and re-initialization of late encoder layers yielded improvement over the standard fine-tuning of BERT on NLP tasks~\cite{zhang2021:revisiting}.
The LLRD strategy was initially developed to tune the later encoder layers with larger learning rate.
In this way, the later layers can be better adapted to a downstream task under consideration, because the later layers are assumed to learn complex task-specific features of input sequences~\cite{howard2018:universal}. 
Moreover, Peters et al.~\cite{peters2019:tune} showed that the performance of fine-tuning improves if the encoder layers are updated during fine-tuning in comparison with training only the classifier on top of fixed (frozen) encoder layers.

Pruning later layers of transformer models is another way to consider only early layers for fine-tuning~\cite{fan2019:reducing, peer2022:greedylayer, sajjad2023:effect}.
Sajjad et al.~\cite{sajjad2023:effect} investigated how the performance of transformer models on NLP is affected when reducing their size by pruning layers.
They considered six pruning strategies, including dropping from different directions, alternated layer dropping, or dropping layers based on importance, for four pre-trained models: BERT~\cite{devlin2019:bert}, RoBERTa~\cite{liu2019:roberta}, XLNET~\cite{yang2020:xlnet}, ALBERT~\cite{lan2020:albert}. By pruning model layers, Sajjad et al. were able to reduce the number of parameters to 60\% of the initial parameter set while maintaining a high level of performance. 
While the performance on downstream tasks varies in their study,
the lower layers are critical for maintaining performance when fine-tuning for downstream tasks. 
In other words, dropping earlier layers is detrimental to performance.
Overall, pruning layers reduces model size and in turn reduces fine-tuning and inference time.
In line with the work of Sajjad et al.~\cite{sajjad2023:effect}, we extend our experiments with the pruning of later layers and keeping earlier layers present in the model (see RQ2 in Section~\ref{sec:results}).

The use of information from single early layers in a number of \method{} experiments is also inspired by Peters et al.~\cite{peters2018:dissecting}. 
In their study, Peters et al. present an empirical evidence that language models learn syntax and part-of-speech information on earlier layers of a neural network, while more complex information, such as semantics and co-reference relationships, are captured better by deeper (later) layers.
In another study, Karmakar and Robbes probed pre-trained models of code, including CodeBERT, on tasks of understanding syntactic information, structure complexity, code length, and semantic information~\cite{karmakar2021:what}. 
While Karmakar and Robbes probed frozen early layers of different models for code in a single strategy, we use 12 different strategies for combining unfrozen early layers during fine-tuning and focus on the tasks of bug detection or bug type classification. 
\maxh{Similarly, Hernández López et al.~\cite{hernandezlopez2023:astprobe} probed different layers of five pre-trained models, including CodeBERT~\cite{feng2020:codebert} and GraphCodeBERT~\cite{guo2021:graphcodebert}, and found that most syntactic information is encoded in the middle layers.}
The novelty of our study with respect to Karmakar and Robbes is that we combine early layers in addition to extracting each of them, while Karmakar and Robbes extracted early layer representations and used them without composing new representations.

\section{Encoders for Code Classification}  %
\label{sec:background}

\ana{In this section, we present the background on transformer \leon{models} and different uses of \leon{the} encoder-decoder---or full transformer--- architecture, \leon{as well as its} encoder-only and decoder-only variants. 
Because our study focuses on encoder-only open-source models available for fine-tuning, the distinction between transformer types is necessary for understanding the methodology.}

In sequence-to-sequence generation scenarios, the transformer model consists of a multi-layer encoder that represents the input sequence and a decoder that generates the output sequence based on the sequence representation from the encoder and the available output generated at previous steps~\cite{vaswani2017:attention}.
For source code classification tasks, the transformer is frequently reduced to only its encoder followed by a \emph{classification head},
a component added to the encoder to categorize the representation into different classes. 
Dropping the decoder for classification is motivated by resource efficiency, because the decoder is conceptually only needed for token generation from the input sequence.
During classification of an input, the encoder represents the sequence and passes it to the classification head.  
Based on this design, a number of pre-trained encoders have been published in recent years, such as BERT and RoBERTa which were pre-trained on natural language, and similar models pre-trained on code, or a combination of code and natural language~\cite{devlin2019:bert, liu2019:roberta}. 
The goal of pre-training in the \emph{pre-train and fine-tune} scenario is to capture language patterns in general, 
so that they can serve as a basis for domain-specific downstream tasks. 
Pre-trained models can be fine-tuned on different downstream tasks in NLP and SE. 

Processing the input sequence for classification consists of several steps: \emph{tokenization, initial embedding, encoding} the sequence with an encoder, and passing the sequence representation through a \emph{classification head}. 
Tokenization splits the input sequence, adds special tokens, matches the tokens to their ID's in the vocabulary of tokens, and unifies the resulting token length for samples in a dataset. 
Embedding transforms the one-dimensional token ID to an initial multi-dimensional static vector representation of the token and is usually a part of the pre-trained encoder model. 
This representation is updated using the attention mechanism of the encoder. 
Because of attention, the representation of the input is influenced by all tokens in the sequence, so it is contextualized. 

CodeBERT is a RoBERTa-based model with 12 encoder layers pre-trained on 6 programming languages (Python, Java, JavaScript, PHP, Ruby, and Go), as well as text-to-code tasks~\cite{feng2020:codebert}. 
Pre-training was done on the masked language modeling (MLM) and replaced token detection (RTD) tasks. 
These tasks respectively train the model to derive what token is masked in MLM, and in RTD predict whether any token in an original sequence is swapped with a different token that should not be in the sequence.  
CodeBERT outputs a bidirectional encoder representation of the input sequence, which means that the model considers context from pre-pending and subsequent words to represent each token in the input sequence.  

A pre-trained model is usually released with a pre-trained tokenizer.
The pre-trained tokenizer ensures that token ID's correspond to those processed during pre-training.
The tokenizer also adds special tokens, such as a \CLS token at the start of each input sequence, \verb#PAD# tokens to unify lengths of input sequences, and the \verb#EOS# token to signify the end of the input string and the start of padding sequence~\cite{devlin2019:bert}. 
All tokens are transformed by the model in each encoder layer. 
Out of all tokens, the \CLS token representation from the last layer, which is updated by all encoder layers, is typically used as a representation for the full sequence. 

The standard practice of using the \CLS token from the last encoder layer is motivated by the pre-training procedure. 
For example, in MLM, the model predicts the masked token based on the \CLS token representation from the 12\textsuperscript{th} layer of BERT and CodeBERT. 
However, the choice of token to represent the full sequence in fine-tuning can be different. 
For example, in PLBART~\cite{ahmad2021:unified}, a transformer model for code with both an encoder and a decoder, the \verb#EOS# token is used for representing the input sequence. 
In this paper, we propose different ways to represent the input sequence and use information from early layers of the model in an effective way.

\section{Methodology}
\label{sec:methodology}

\begin{figure}[b]
\centering
\includegraphics[width=\columnwidth, trim={0mm 5mm 0mm 5mm}]{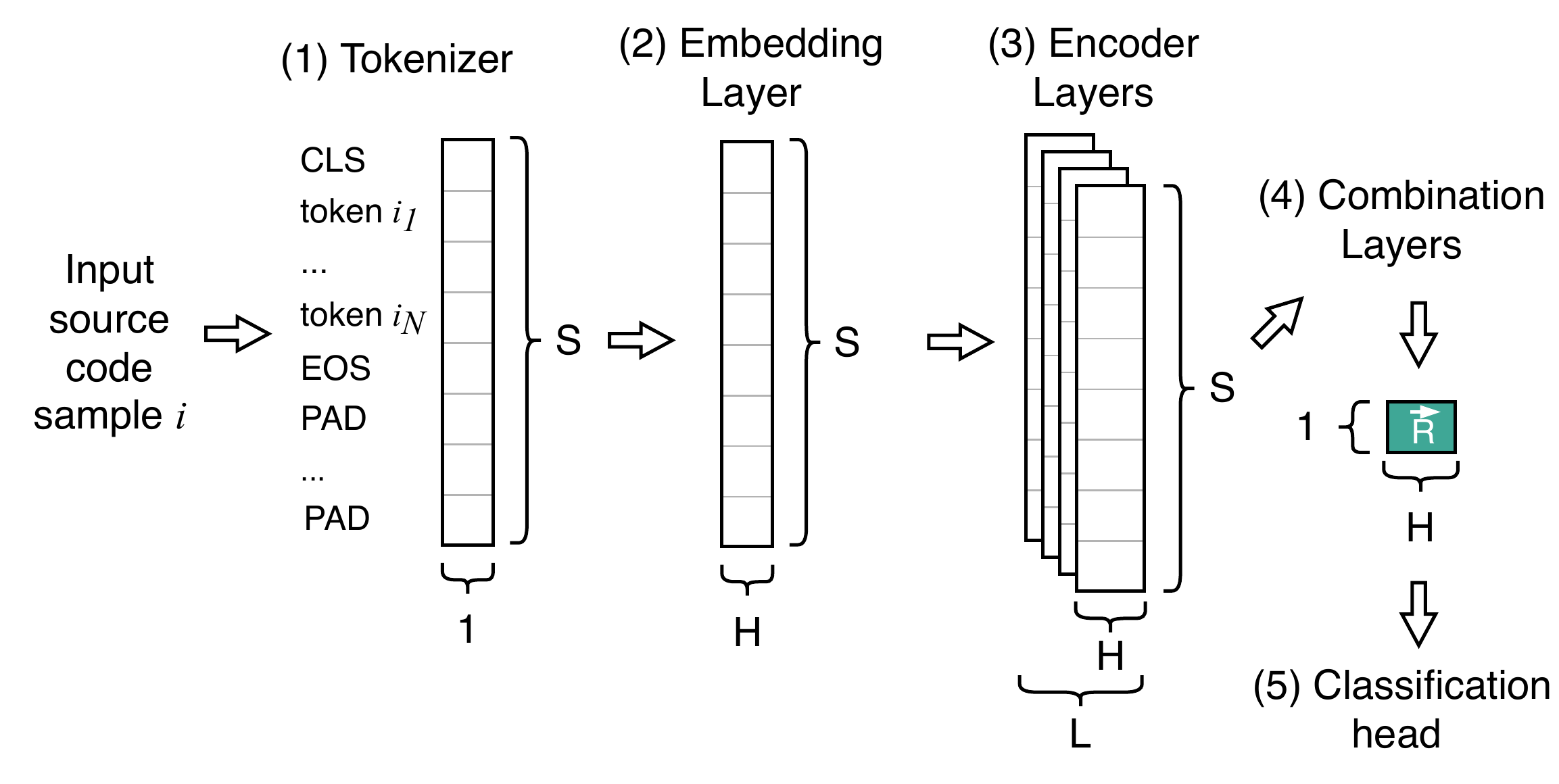}
\caption{Model architecture for code classification.}
\label{fig:method-overview}
\end{figure}

In this paper, the architecture of the code classification model consists of five parts: (1) a tokenizer, (2) an embedding layer, (3) an encoder with several layers, (4) a set of operations to combine sequence representations from encoder layers with \method{}, and (5) a classification head. 
The output of each step is used as input into the next step. 
An overview of the architecture is shown in Figure~\ref{fig:method-overview} and described below. 
The main difference between this architecture and the classification architecture discussed in Section~\ref{sec:background} is step (4);
the standard architecture only consists of steps (1--3) and (5).

Steps (1)--(3) use a pre-trained tokenizer, embedder, and encoder. 
\method{} is formulated in a generic way and can be applied to any encoder, but for our experiments, we fix the CodeBERT model and tokenizer. 
In step~(4), we combine information from all the layers or from only some of the early layers
of the encoder, as opposed to the baseline that uses the last layer of the encoder. 
Finally, the classification head in step (5) consists of one dropout layer and one linear layer with softmax. 

The encoder model represents each token of an input sequence with a vector of size $H$, also known as hidden size.
For each input sequence of length $S$, and a hidden size $H$, we obtain a matrix of size $S \times H$ for each of $L$ layers of the base model as shown in Figure~\ref{fig:method-overview}.  
For example, the CodeBERT architecture is fixed with 12 encoder layers, i.e., $L = 12$ for that model.
All the information available in the encoder for one input sequence is stored in a tensor of size $L \times S \times H$. 
The \method{} combinations must produce one vector $\vec{R}$ of size $H$ that represents the input, as shown in Figure~\ref{fig:method-overview}. 
Keeping the output code representation of size $H$ is required to provide a fair comparison of \method{} composite representations with the standard code representation obtained from the last layer. 
In this way, the dimension of the classification head is the same for all combinations of early layers and has minimal possible influence during fine-tuning.

\ana{\leon{As a strategy for systematically investigating composite representations, we} create a grid-search over three typical operations to combine outputs of neural network layers -- maximum pooling (max pool), weighted sum and slicing -- and two dimensions to apply the operations: over tokens and/or layers. 
For the tokens dimension, we either use all of the tokens from a specific layer or only the \CLS token. 
Among layers, we either slice one layer, sum or take maximum values over all layers.
}
\maxh{The choice of considering every token of a layer is motivated by the fact the transformer-based models exhibit varying degrees of attention for different types of tokens~\cite{paltenghi2021:thinking}, which indicates that solely using the \CLS token might not be the best choice for tasks~\cite{sharma2022:exploratory}.}
We also experiment with different sizes of the model. 
The combination strategies that use all layers of the pre-trained model are divided into two categories: the strategies that use \CLS tokens from the encoder layers; the strategies that use more tokens than just {\CLS} 
 from encoder layers. 

\ana{When we slice the \CLS token and apply each of the operations over layers, we obtain the following \CLS-token combinations:}
\begin{enumerate}[label=(\roman*)]
    \item \emph{baseline}: \CLS token from the last layer, i.e., layer no. $L$; 
    \item \CLS token from one layer\footnote{~\ana{We use each layer $l$ in the combinations separately if we denote $l, \; l \in \{1, \dots, L\},$ and specify the set of layers 
    $\{l\}_{l=1}^L$ 
    if several layers are used at once.}} no. 
    $l, \; l \in \{1, \dots, (L-1)\};$
    \item max pool over \CLS tokens from all layers 
    $\{l\}_{l=1}^L$;%
    \item weighted sum over \CLS tokens from all layers 
    $\{l\}_{l=1}^L$.
\end{enumerate}

\begin{figure*}

\vspace*{1ex}
\begin{subfigure}{.25\textwidth}
\centering
\includegraphics[width=0.7\textwidth, trim={5mm 0mm 5mm 0mm}]{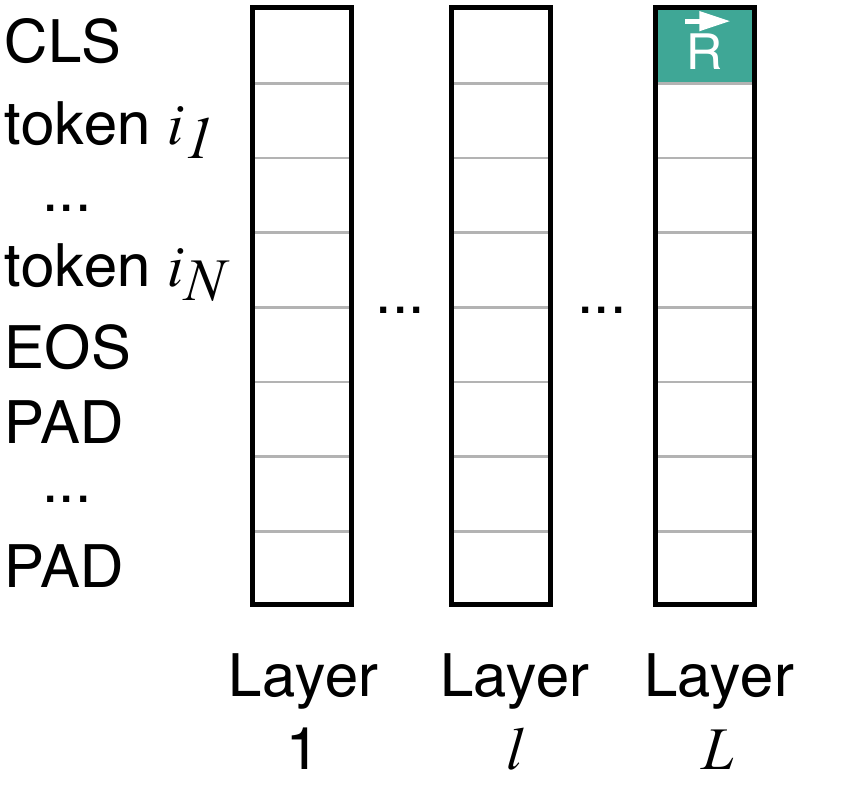}
\caption{\emph{Baseline}: CLS token of layer L (i).}
\label{fig:all-comb-a}
\end{subfigure}
\hspace*{20pt}
\begin{subfigure}{.25\textwidth}
\centering
\includegraphics[width=0.692\textwidth, trim={5mm 0mm 5mm 0mm}]{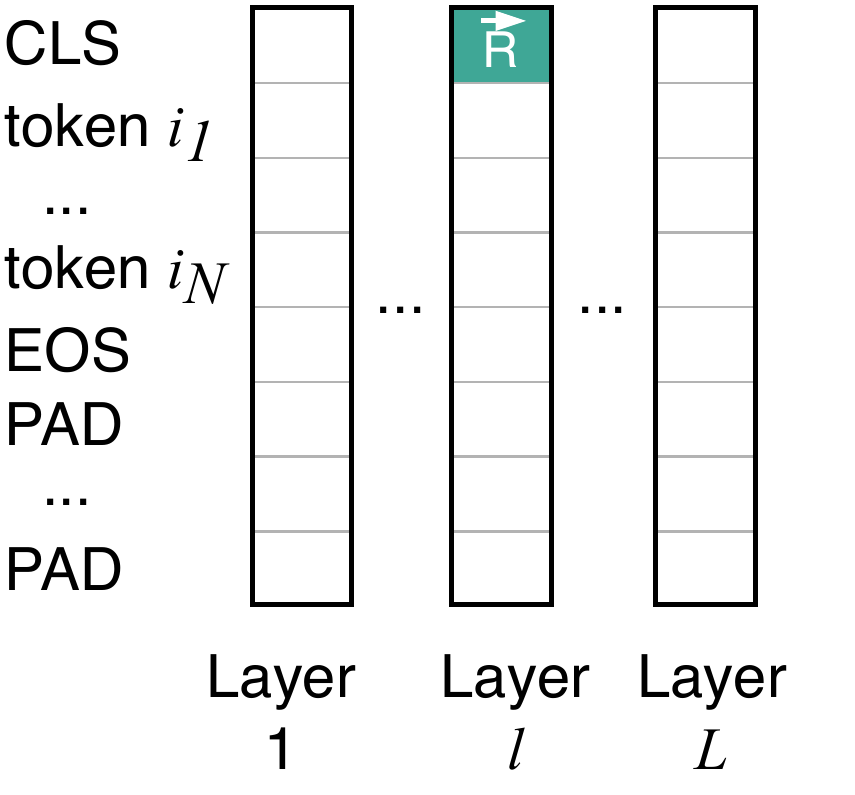}
\caption{Full model, CLS of layer \textit{l} < L (ii).}
\label{fig:all-comb-b}
\end{subfigure}
\hspace*{20pt}
\begin{subfigure}{.3\textwidth}
\centering
\includegraphics[width=0.427\textwidth, trim={5mm 0mm 5mm 0mm}]{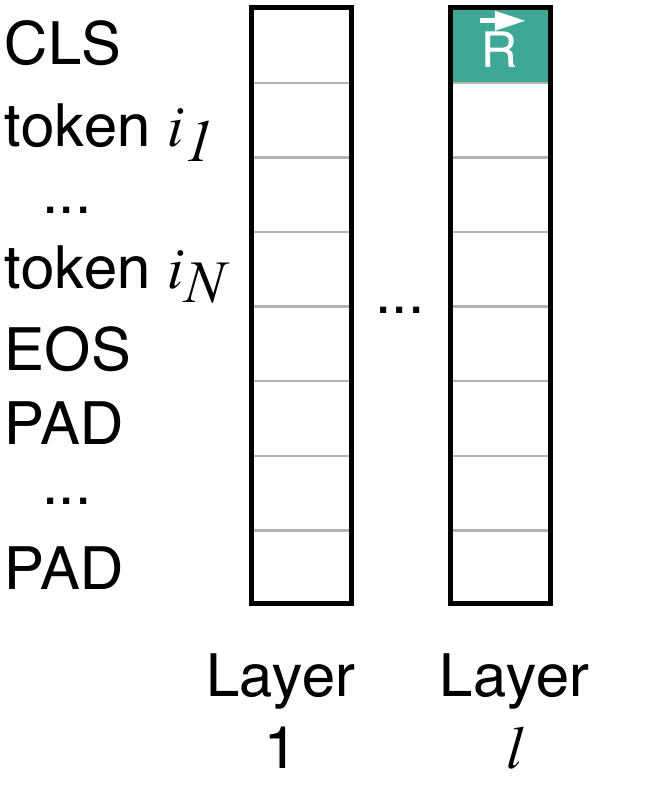}
\caption{Pruned model, CLS of last layer \textit{l}~(xii).} 
\label{fig:all-comb-c}
\end{subfigure}
\vspace{4ex}

\begin{subfigure}{0.9\columnwidth}
\centering
\includegraphics[width=.95\textwidth, trim={0mm 0 0mm 0}]{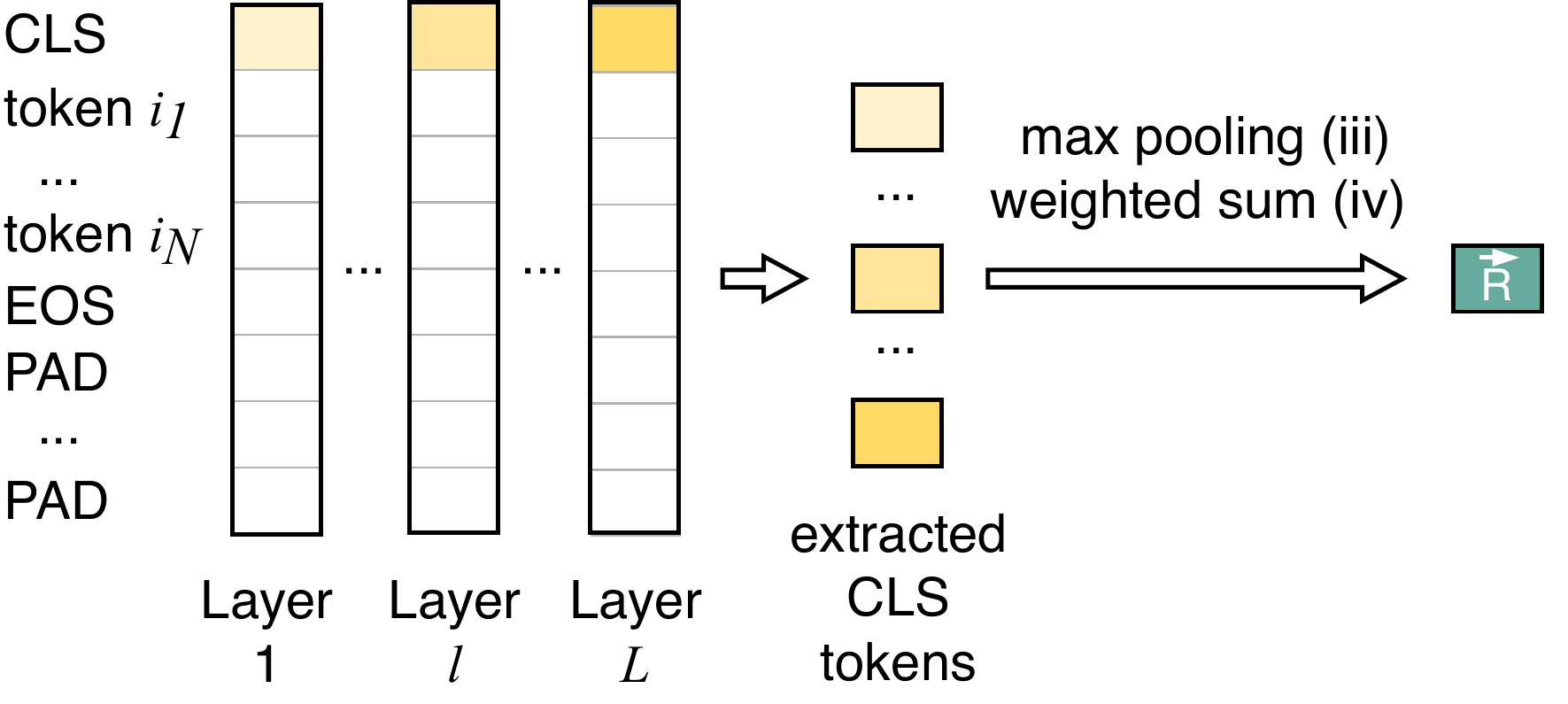}
\caption{Combinations of CLS tokens from all layers (iii, iv).}
\label{fig:all-comb-d}
\end{subfigure}
\hspace{25pt}
\begin{subfigure}{.9\columnwidth}
\centering
\includegraphics[width=.945\textwidth, trim={0 0 0 0}]{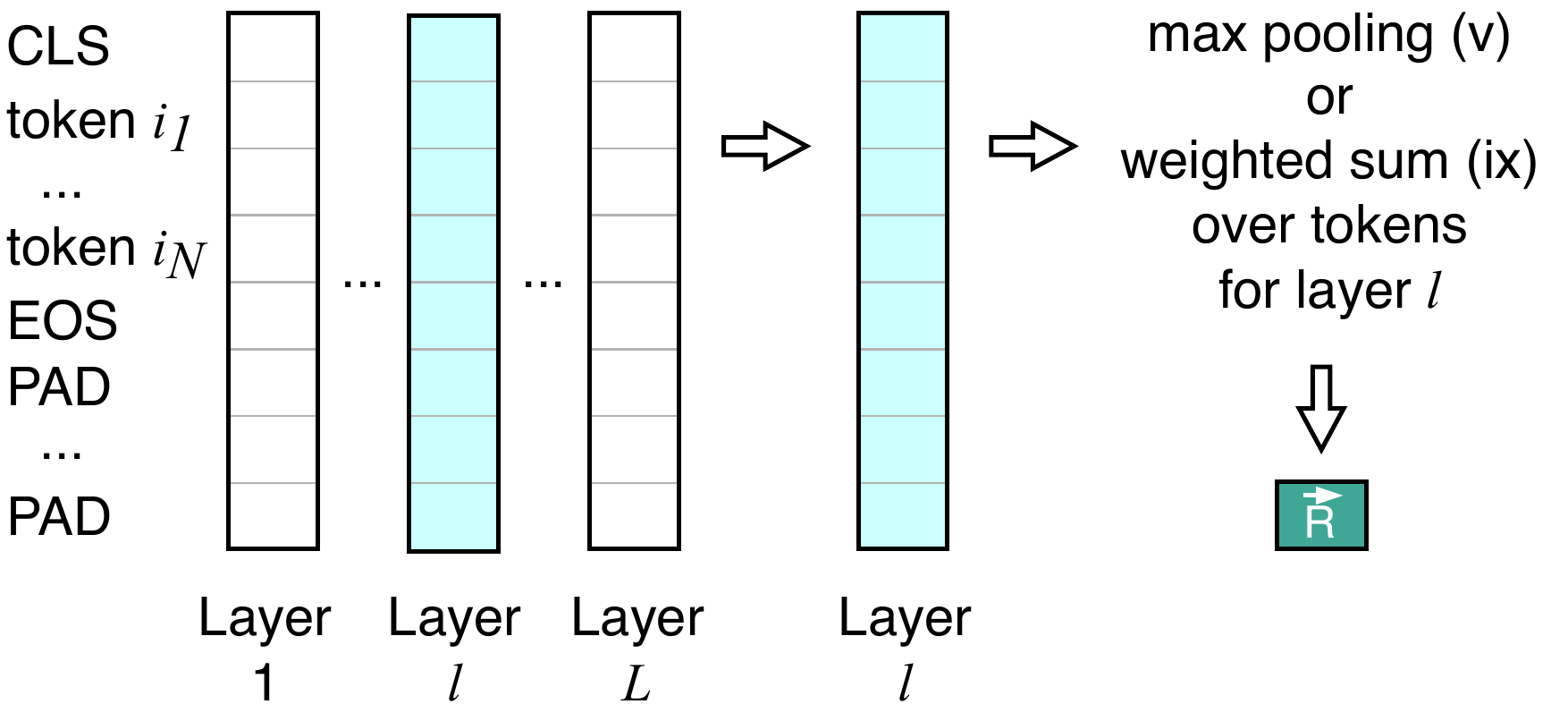}
\caption{Combinations of all tokens from a single layer (v, ix).}
\label{fig:all-comb-e}
\end{subfigure}
\vspace{4ex}

\begin{subfigure}{\columnwidth}
\centering
\includegraphics[width=\textwidth, trim={0 0 0 0}]{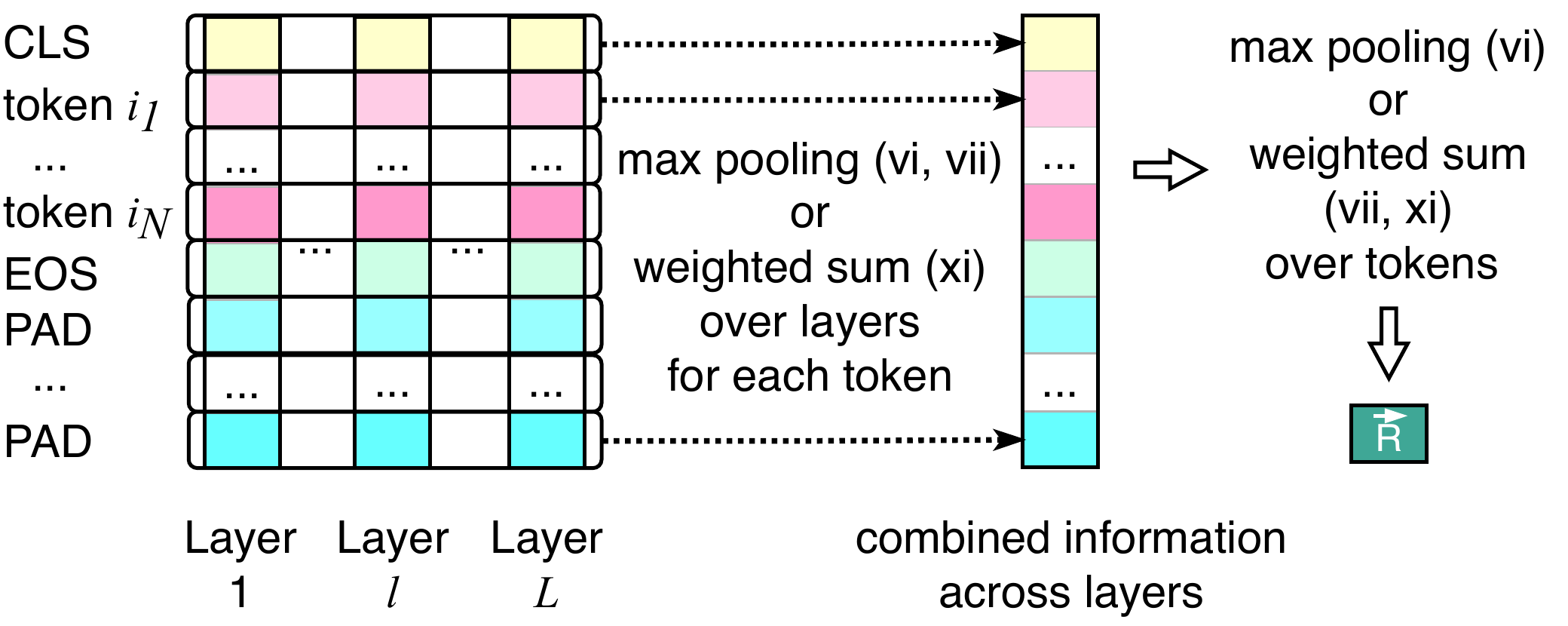}
\caption{Combinations of all tokens combined across layers using max pooling or weighted sum (vi, vii, xi).}
\label{fig:all-comb-f}
\end{subfigure}
\hspace{15pt}
\begin{subfigure}{\columnwidth}
\centering
\includegraphics[width=\textwidth, trim={4mm 0 4mm 0}]{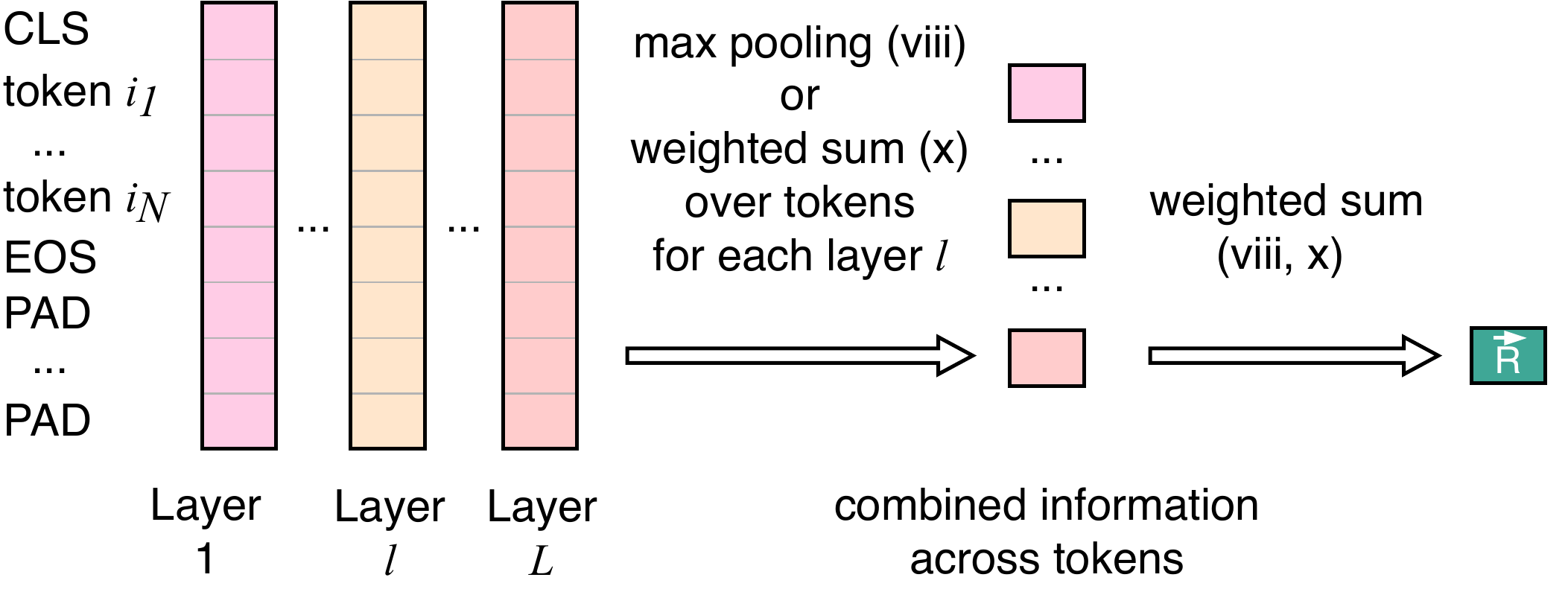}
\caption{Combinations of all tokens combined for each single layer first using max pooling or weighted sum (viii, x).}
\label{fig:all-comb-g}
\end{subfigure}

\caption{Combinations of early encoder layers that lead to code representation vector $\vec{R}$ for each tokenized input sequence. 
The latin numbering in brackets corresponds to the combinations described in Section~\ref{sec:methodology}. 
\leon{Observe that the presentation order has been designed to preserve space by grouping similar combinations in the same subfigure.}
}
\label{fig:all-comb}
\end{figure*}

The second set of combinations uses representations of all the tokens in tokenized input sequences, including the \CLS token. 
\ana{We first apply max pooling operation to either all tokens or all layers and use the rest of operations. 
Then we apply weighted sum as the first operation followed by max pool or slicing of a layer:}
\begin{enumerate}[label=(\roman*)]
\setcounter{enumi}{4}
    \item max pool tokens from one layer no. $l, \; l \in \{1, \dots, L\};$
    \item max pool over all layers for each token in the input sequence, max pool over tokens;
    \item max pool over all layers for each token in the input sequence; weighted sum over tokens;
    \item max pool over all tokens for each layer no.  $l, \; l \in \{1, \dots, L\};$ weighted sum over layers
    \item weighted sum over tokens from one layer no. $l, \; l \in \{1, \dots, L\};$
    \item weighted sum over tokens for each one layer no. $l, \; l \in \{1, \dots, L\};$ weighted sum over all layers; 
    \item weighted sum over all layers for each token in the input sequence; weighted sum over all tokens.
\end{enumerate}
Note that weights in the weighted sums are learnable parameters. 
However, the added number of learnable parameters for fine-tuning constitutes 0.00042\%\footnote{\ana{~Weighted sum over tokens adds $S = 512$ learnable weights. Because the weights of the sum are shared across the layers, the maximum number of added weights is $L + H = 524$ out of 124M learnable weights in the base model. Combinations without weighted sums do not add extra learnable parameters to the base model. Weighted sum over layers adds learnable $L=12$ weights for CodeBERT.}} of the number of learnable parameters in the baseline configuration. 
For this reason, we mention that the models with combinations (ii-x) have the same model size while bearing in mind the overhead of learnable weights in the weighted sums. 

In addition to experiments with token combinations, we also investigate performance of the model with first $l<L$ layers and the baseline token combination, described as follows: 
\begin{enumerate}[label=(\roman*)]
\setcounter{enumi}{11}
    \item \CLS token from the last layer of the model with $l<L$ encoder layers.
\end{enumerate}
Note that the baseline combination (i) with the usage of the \CLS token from layer $L$ corresponds to (ii) and (xii) if $l=L.$

The combinations are presented in Figure~\ref{fig:all-comb}. 
Similar combinations are presented close to each other or are combined in the same image if they only have minor differences and share the major parts. 
For example, in Figure~\ref{fig:all-comb-c}, we illustrate combinations (iii) and (iv), because both of them use \CLS tokens from all layers combined using max pooling or weighted sum.
The roman numbers which indicate combination types are preserved either in the descriptions below the figures or in the figures themselves, but the order is changed.
We mention combination number corresponding to the description in the current section, 
such as baseline combination (i) in Figure~\ref{fig:all-comb-a} or combination (ii) for \CLS token from one early layer in Figure~\ref{fig:all-comb-b}.
We highlight what parts of encoder layer outputs are used for each combination with color.
White cells correspond to the tokens that are not used in early layer combinations.  
The goal of all combinations is to obtain a vector representation $\vec{R}$ for each input code sample. 
For example, in Figure~\ref{fig:all-comb-a}, we consider the last layer $L$ and extract only the \CLS token marked as $\vec{R}$.

Another remark on the \method{} combinations  concerns the usage of all tokens or only code tokens. 
Code tokens are those that correspond to tokenized input words or sub-words and are shown in Figure~\ref{fig:all-comb} as $\text{token}_{i_1}, ..., \text{token}_{i_N}$ for an input sequence $i$ of size $i_N.$ 
For each combination that uses more than just a \CLS token, i.e., combinations (v-xi), we experiment with code tokens only, as well as with all tokens, including \CLS, \verb#EOS#, and \verb#PAD#. 
The motivation to check code tokens exclusively stems from the hypothesis that information in special tokens may introduce noise into results. 

\section{Experimental Setup}
\label{sec:experiment}
In this section, we describe the datasets used for empirical evaluation and implementation details of fine-tuning with the proposed \method{} approach.
We investigate binary and multi-task code classification scenarios to explore generalisability of our results. 

\subsection{Datasets for Source Code Classification}
\label{sec:datasets}

We fine-tune and test the CodeBERT model using the \method{} approach on four datasets. 
The datasets span three tasks: defect detection, error type classification and exception type classification --- with 2, 3, and 20 classes, respectively. 
They also contain data in two programming languages, C++ and Python.
In addition, the chosen datasets have similar train subset sizes.
In this way, we aim to reduce the effect of the model's exposure to different amounts of training data during fine-tuning. 
Statistics of the datasets are provided in Table~\ref{tab:datasets}.
We report the size of the train/validation/test splits. 
In addition, we compute the average number of tokens in the input sequences upon tokenization with the pre-trained CodeBERT tokenizer. 
Because the maximum input sequence size for the CodeBERT model is limited to $S=512,$ the number of tokens is indicative of how much information the model gets access to or how much information is cut off, in case of long inputs.

\head{Devign} 
This dataset contains functions in C/C++ from two open-source projects labelled as vulnerable or non-vulnerable~\cite{zhou2019:devign}. 
We reuse the train/validation/test split from the CodeXGLUE Defect detection benchmark.\footnote{~\url{https://github.com/microsoft/CodeXGLUE/tree/main/Code-Code/Defect-detection}}
The dataset is balanced: the ratio of non-vulnerable functions is 54\%.

\head{ReVeal} 
Similarly to Devign, ReVeal is a vulnerability detection dataset of C/C++ functions~\cite{chakraborty2021:deep}. 
The dataset is not balanced: it contains 90\% non-vulnerable code snippets.
Both the Devign and ReVeal datasets contain real-world vulnerable and non-vulnerable functions from open-source projects.

\head{Break-It-Fix-It (BIFI)}
The dataset contains function-level code snippets in Python with syntax errors~\cite{yasunaga2021:breakitfixit}. 
We use the original buggy functions and formulate a task of classifying the code into three classes: Unbalanced Parentheses with 43\% of the total number of code examples in BIFI, Indentation Error with 31\% 
code samples, 
Invalid Syntax containing 26\% samples. 
The train/test split provided in the dataset is reused, and the validation set is extracted as 10\% of training data.

\head{Exception Type}
The dataset consists of short functions in Python with an inserted \verb#__HOLE__# token in place of one exception in code.\footnote{~\url{https://github.com/google-research/google-research/tree/master/cubert}} 
The task is to predict one of 20 masked exception types for each input function and is unbalanced.
The dataset was initially created from the ETH Py150 Open corpus\footnote{~\url{https://www.sri.inf.ethz.ch/py150}} as described in the original paper~\cite{kanade2020:learning}.
We reuse the train/validation/test split provided by the authors.

\begin{table}[t]
\caption{Statistics of Fine-Tuning Datasets.}
\label{tab:datasets}
\begin{tabular}{l@{}rrrrr}
\toprule
\multirow{2}{*}{Dataset}  & \multirow{2}{*}{\# classes}  & 
\multirow{2}{3.2em}{\raggedleft Avg \# tokens}      
& \multicolumn{3}{c}{\# code samples} \\
\cmidrule{4-6}
& & & Train & Valid & Test \\ 
\midrule
Devign          & 2         & 614               & 21,854     & 2,732          & 2,732  \\
ReVeal          & 2         & 512               & 18,187     & 2,273          & 2,274  \\
BIFI            & 3         & 119               & 20,325     & 2,259          & 15,055 \\
Exception Type  & 20        & 404               & 18,480     & 2,088          & 10,348 \\
\bottomrule
\end{tabular}
\end{table}

\subsection{Implementation}

The architecture is based on the CodeBERT\footnote{~\url{https://huggingface.co/microsoft/codebert-base}} tokenizer and encoder model.
The model defines the maximum sequence length, hidden size, and has 12 layers, so $S=512$, $H=768$, $L=12.$
Hyper-parameters in the experiments are set to $B=64$, learning rate is $1e\text{-}5$, and dropout probability is $0.1$. 
If the tokenized input sample is longer than $S=512$, we prune the tokens in the end to make the input fit into the model.
We run fine-tuning with Adam optimizer and testing for each combination 10 times with different seeds for 10 epochs and report the performance for the best epoch on average over 10 runs. 
The best epoch is defined by measuring accuracy on a validation set. 
We use Python~3.7 and Cuda~11.6, and run experiments on one Nvidia Volta A100 GPU.

\subsection{Evaluation Metrics}

To present the impact of early layer combinations, we compare the accuracy on the test set for all datasets, because it allows us to compare our results with other benchmarks. 
In addition, we report weighted F1-score denoted as F1(w) for a detailed analysis of selected combinations to account for class imbalance.
To obtain the weighted F1-score, the regular F1-score is calculated for each label and their weighted mean is taken. 
The weights are equal to the number of samples in a class. 

We also report results of the Wilcoxon signed-rank test on the corresponding metrics for the combinations that show improvement over the baseline~\cite{wilcoxon1992:individual}. 
The Wilcoxon test is a non-parametric test suitable for the setting in which different model variants are tested on the same test set, because it is a paired test. 
The Wilcoxon test checks the null hypothesis whether two related paired samples come from the same distribution. 
We reject the null hypothesis if p-value is less than $\alpha=0.05.$ 
In case we obtain improvement of a metric over the baseline with an \method{} combination and the null hypothesis is rejected, we conclude that the combination performs better and the result is statistically significant. 
\ana{For the pruned models, we compute Vargha and Delaney's ${A}_{12}$ non-parametric effect size measure of the performance change for accuracy and F1(w) with thresholds of 0.71, 0.64 and 0.56 for large, medium and small effect sizes~\cite{vargha2000:critique}.}

\subsection{Research Questions}
During our empirical evaluation of composite \method{} code representations, we address the following research questions:

\medskip
\head{RQ1. Composite Code Representations with Same Model Size} What is the effect of using combinations (ii-xi) of early layers with the same model size in comparison to the baseline approach of using only the \CLS token from the last layer, i.e., combination (i), for code representation on model performance in the code classification scenario? The goal is to find out whether any of the \method{} combination types work consistently better for different datasets and tasks. 

\medskip
\head{RQ2. Pruned Models}  
What is the effect of reducing the number of pre-trained encoder layers in combinations (xii) on resource usage and model performance on code classification tasks?
As opposed to RQ1, in which we consider the combinations that do not reduce the model size, this research question is devoted to investigation of the trade-off between using less resources with reduced-size models and performance variation in terms of classification metrics. 

\medskip
For both research questions, we evaluate the composite representations on binary and multi-task code classification scenarios to explore generalisability of the results obtained for the binary case. 
We investigate if and what combinations result in better performance, averaged over 10 runs with different seeds. 
For combinations that improve the baseline on average, we also explore if the results are statistically significant according to the Wilcoxon test.

\section{Results and Discussion}
\label{sec:results}

\subsection{\method{} with Fixed-Size Models}
To answer RQ1, we explore one-layer combinations, multi-layer combinations, and estimate the statistical significance of the performance improvement.

\subsubsection{Combinations of Tokens in Single Selected Early Layers}
\label{sec:one-layer}

\begin{figure}[t]
\begin{subfigure}[t]{\columnwidth}
\centering
\includegraphics[width=\textwidth, trim={0.3cm .3cm 0.3cm 0.9cm}, clip]{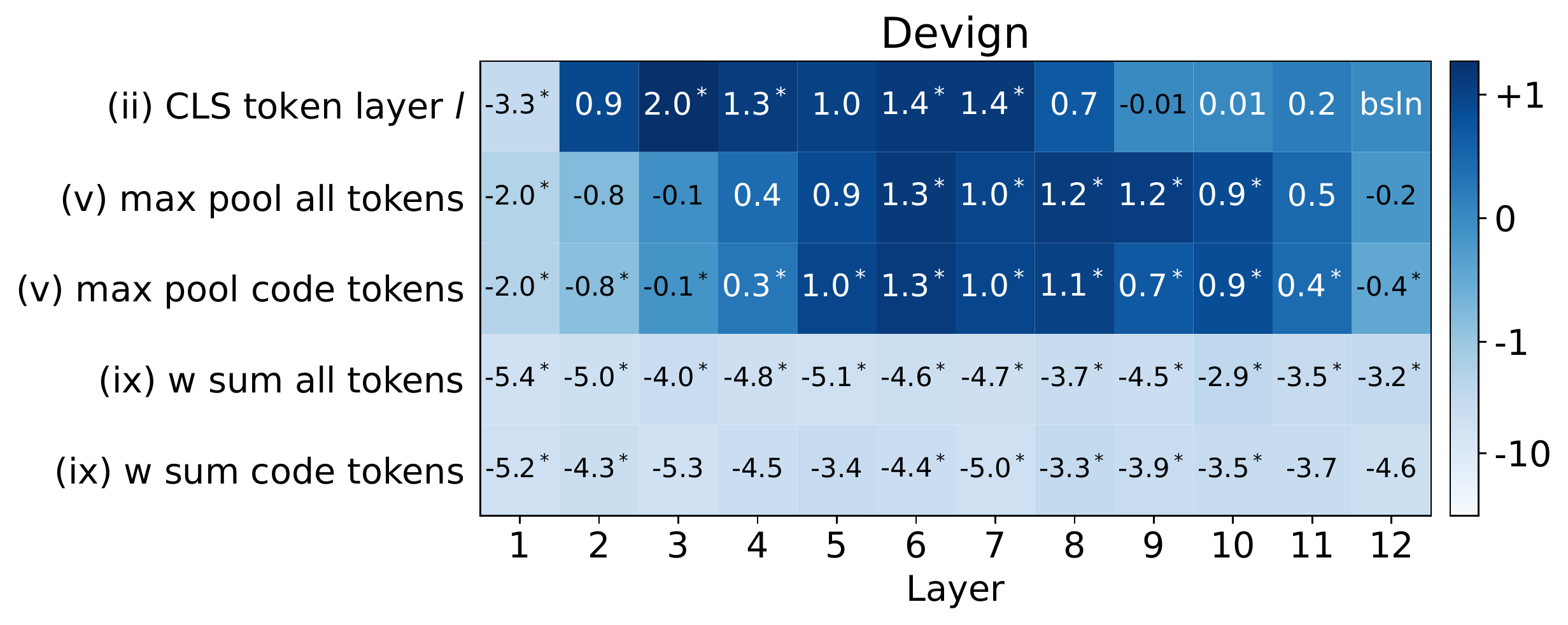}
\caption{Devign, accuracy.}
\label{fig:devign-one-layer-heatmap}
\end{subfigure}
\vspace{1ex}

\begin{subfigure}[t]{\columnwidth}
\centering
\includegraphics[width=\textwidth, trim={.3cm .3cm 0.3cm 0.9cm}, clip]{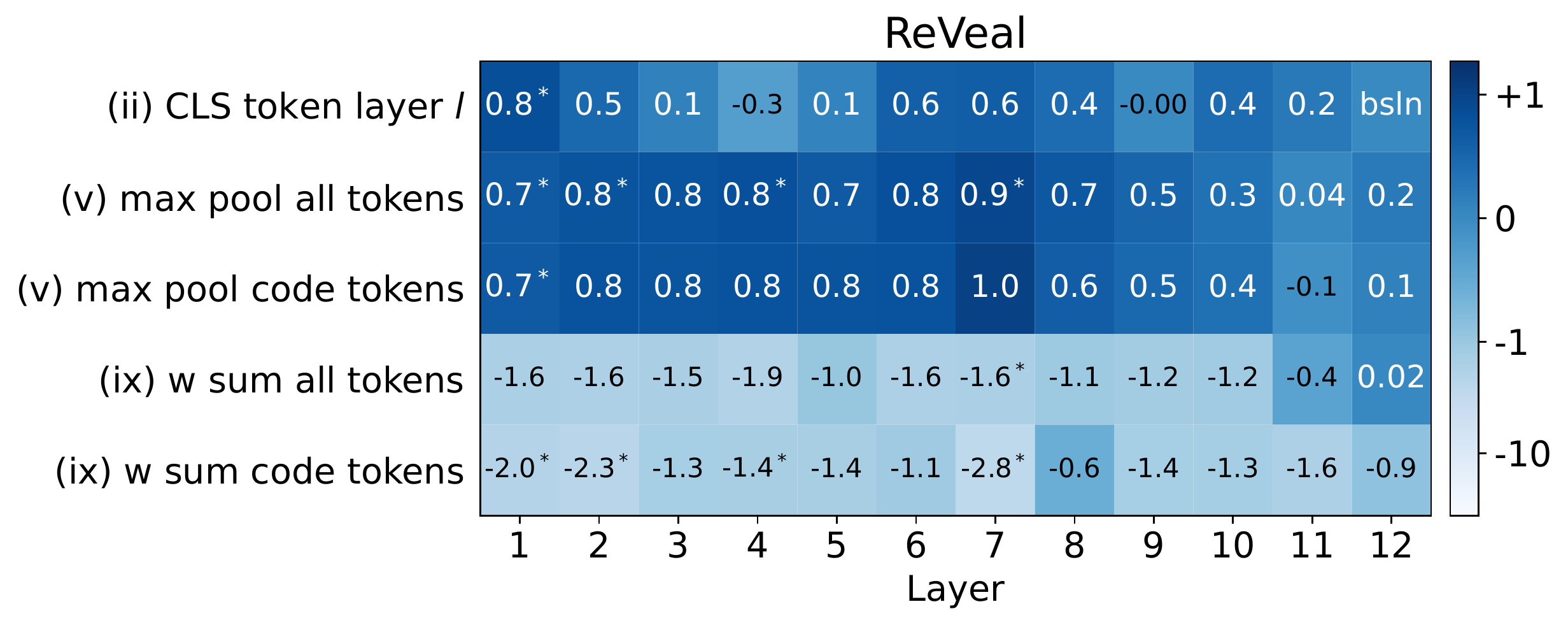}
\caption{ReVeal, accuracy.}
\label{fig:reveal-one-layer-heatmap}
\end{subfigure}
\vspace{1ex}

\begin{subfigure}[t]{\columnwidth}
\centering
\includegraphics[width=\textwidth, trim={0.3cm .3cm 0.3cm 0.9cm}, clip]{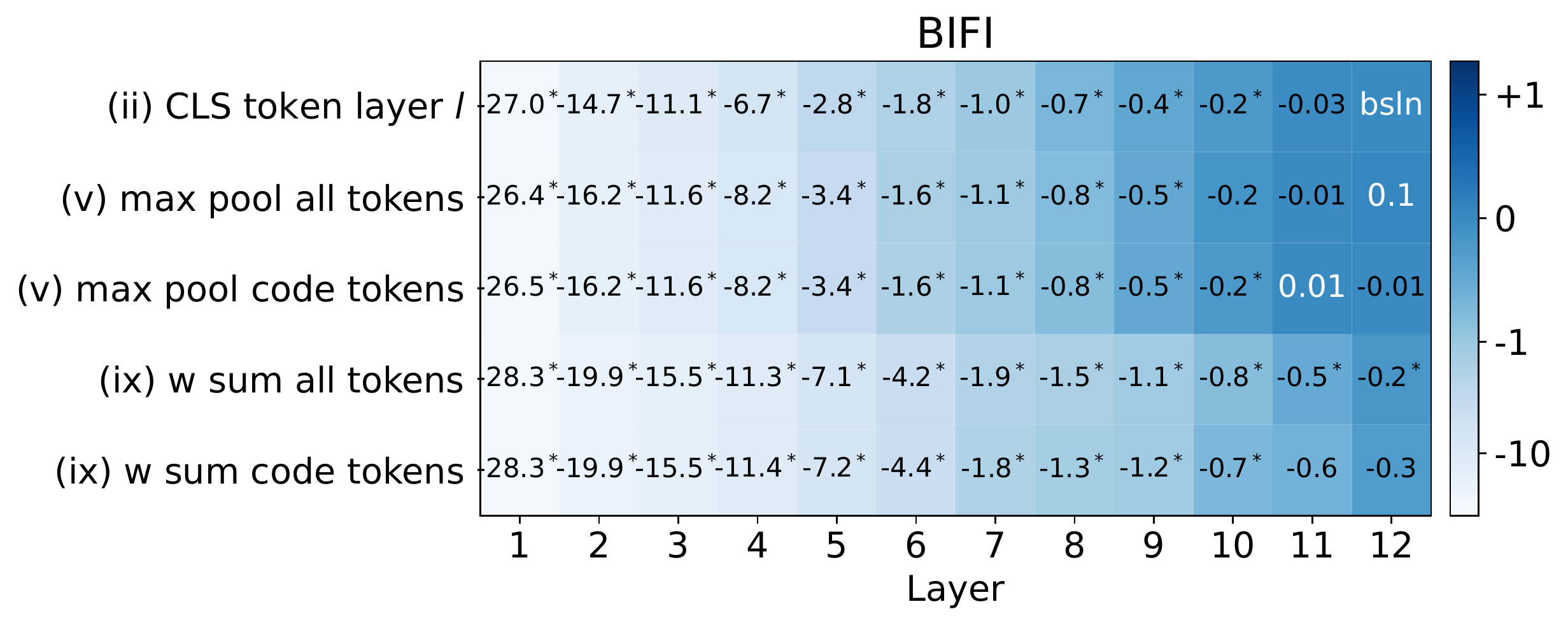}
\caption{BIFI, accuracy.}
\label{fig:bifi-one-layer-heatmap}
\end{subfigure}
\vspace{1ex}

\begin{subfigure}[t]{\columnwidth}
\centering
\includegraphics[width=\textwidth, trim={0.3cm .3cm 0.3cm 0.9cm}, clip]{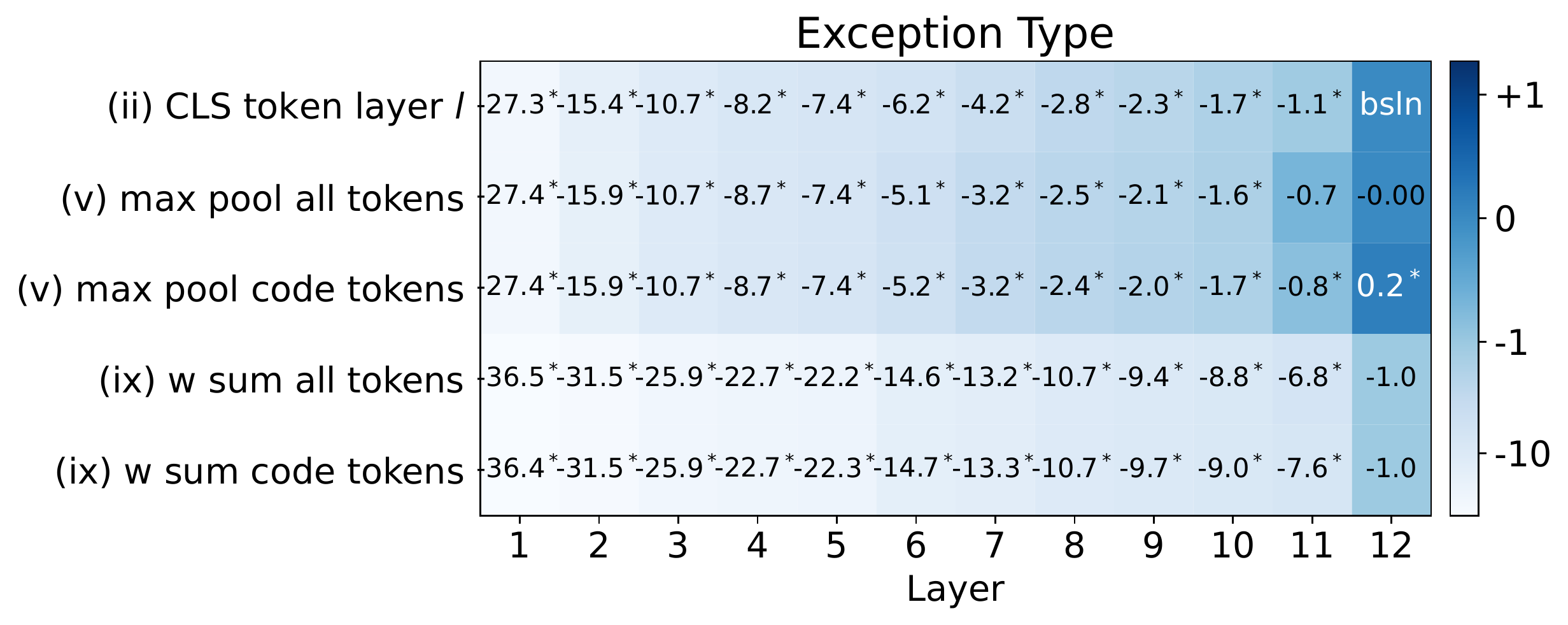}
\caption{Exception Type, accuracy.}
\label{fig:exception-one-layer-heatmap}
\end{subfigure}

\caption{Difference of mean accuracy between \method{} and baseline (\textsf{bsln}) performance. 
The star~$^*$ indicates a statistically significant difference w.r.t. the baseline.
}
\label{fig:one-layer-heatmap}
\end{figure}

\begin{figure}[t] 
\begin{subfigure}[t]{\columnwidth}
\centering
\includegraphics[width=\textwidth, trim={0.3cm 0.3cm 0.3cm 0.9cm}, clip]{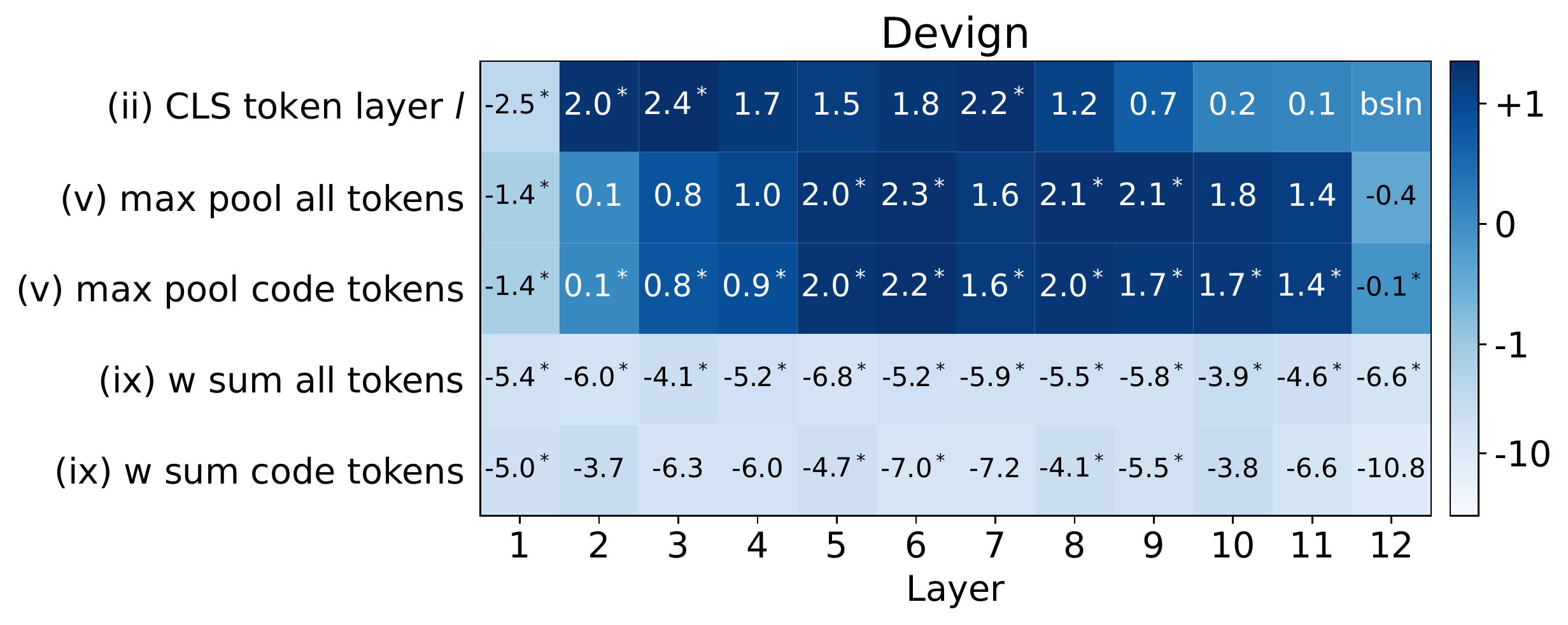}
\caption{Devign, F1(w).}
\label{fig:devign-one-layer-heatmap-f1}
\end{subfigure}
\vspace{1ex}

\begin{subfigure}[t]{\columnwidth}
\centering
\includegraphics[width=\textwidth, trim={0.3cm 0.3cm 0.3cm .9cm}, clip]{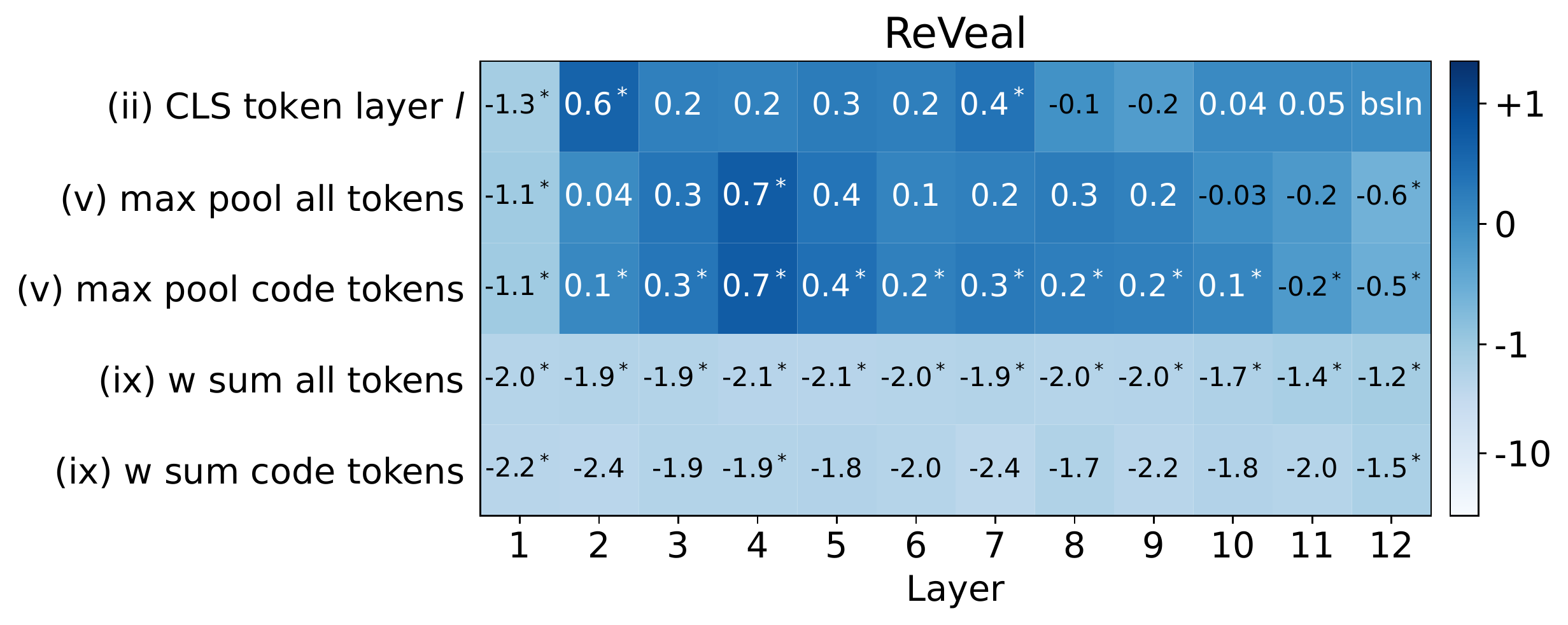}
\caption{ReVeal, F1(w).}
\label{fig:reveal-one-layer-heatmap-f1}
\end{subfigure}
\vspace{1ex}

\begin{subfigure}[t]{\columnwidth}
\centering
\includegraphics[width=\textwidth, trim={0.3cm 0.3cm 0.3cm .9cm}, clip]{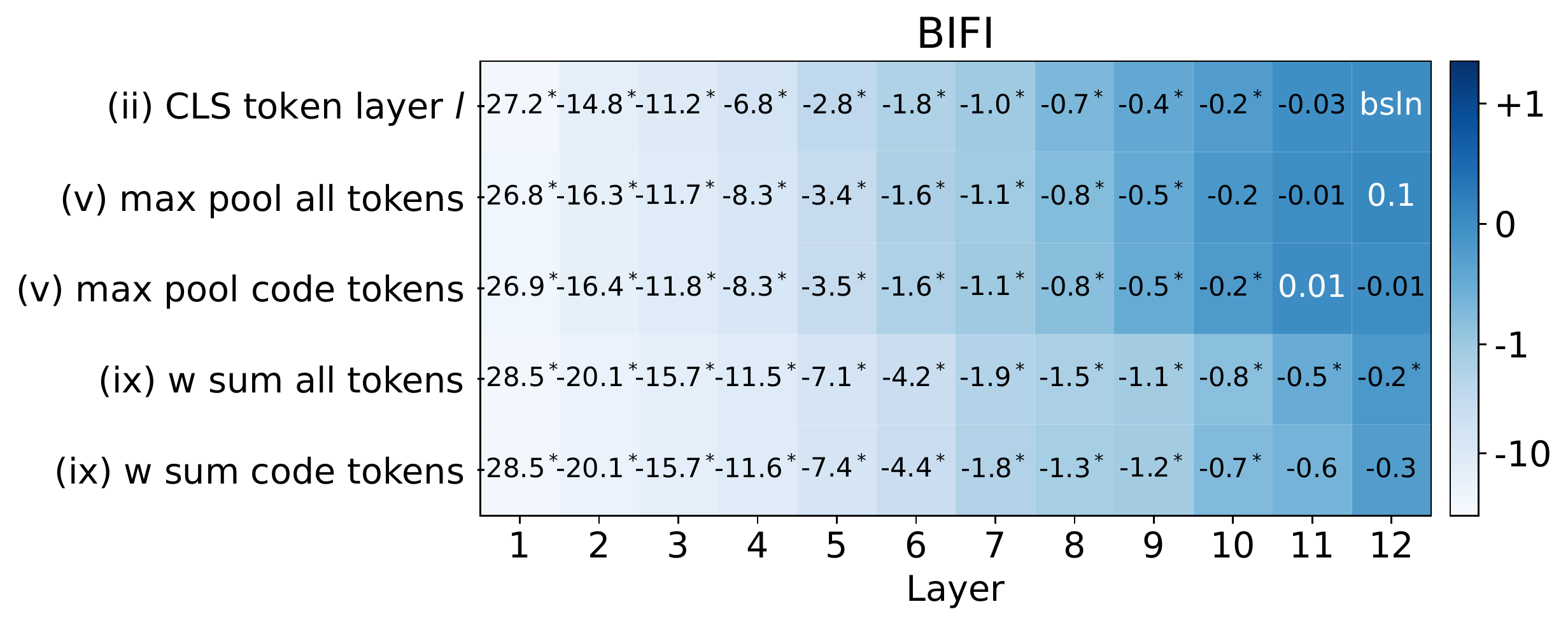}
\caption{BIFI, F1(w).}
\label{fig:bifi-one-layer-heatmap-f1}
\end{subfigure}
\vspace{1ex}

\begin{subfigure}[t]{\columnwidth}
\centering
\includegraphics[width=\textwidth, trim={0.3cm 0.3cm 0.3cm .9cm}, clip]{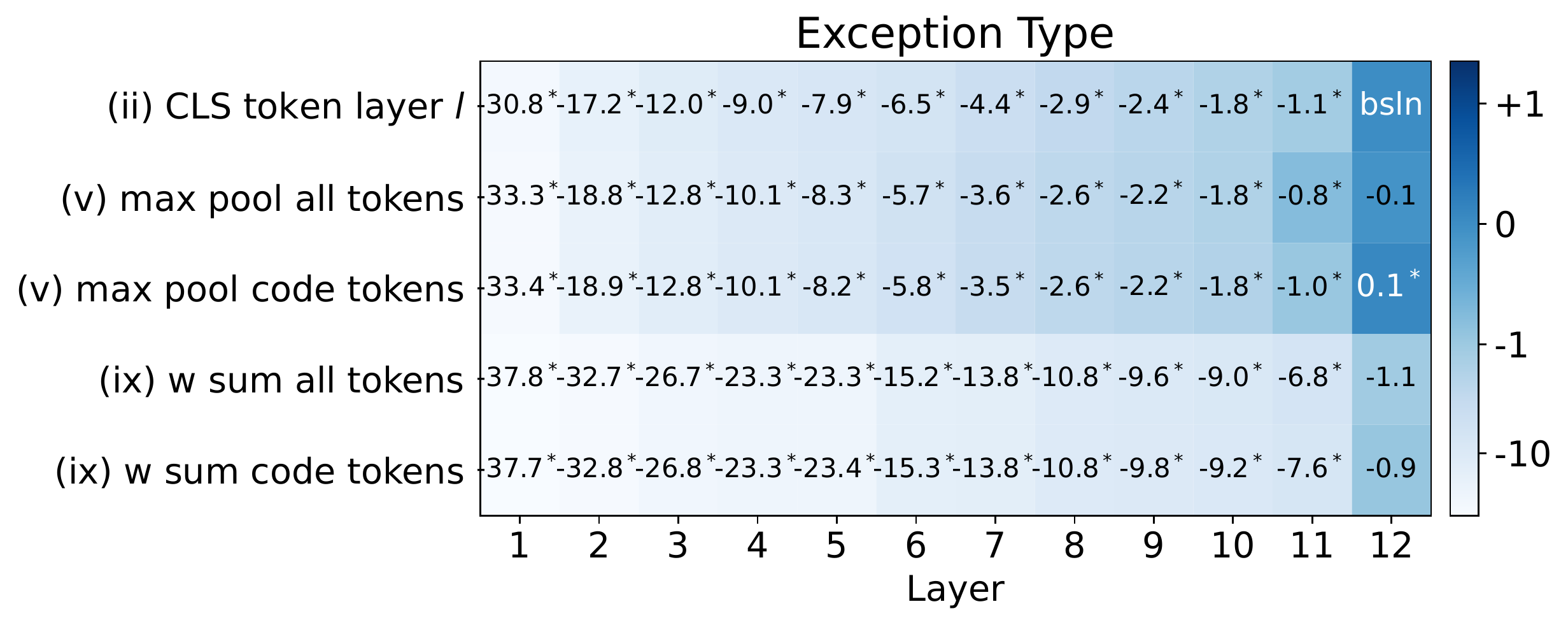}
\caption{Exception Type, F1(w).}
\label{fig:exception-one-layer-heatmap-f1}
\end{subfigure}

\caption{Difference of mean weighted F1-scores (F1(w)) between \method{} and baseline (\textsf{bsln}). 
The star~$^*$ indicates a statistically significant difference w.r.t. the baseline.
}
\label{fig:one-layer-heatmap-f1}
\end{figure}

Figure~\ref{fig:one-layer-heatmap} shows a heatmap of the difference of the mean accuracy obtained with each combination that uses only one selected early layer \ana{compared to the baseline}.
In addition, we show the value of the 
\ana{difference in mean accuracy} 
for each combination type and layer number.
\leon{Note that the scale is logarithmic and in the most extreme case spans the interval from ca.~-37 to +2.}
\leon{Negative values are shown in black, and positive values are shown in white.}
Differences that are statistically significant according to the Wilcoxon test are marked with a star ($^*$) next to the value.
Combinations that correspond to the baseline are marked with ``bsln'' and have zero difference, by definition. 
The results for \leon{the} weighted F1-score \leon{show a similar pattern} as those for \leon{the} mean accuracy. 
They are visualized in the same way in Figure~\ref{fig:one-layer-heatmap-f1}.

The first rows in Figures~\ref{fig:devign-one-layer-heatmap} and~\ref{fig:reveal-one-layer-heatmap} correspond to the combinations~(ii) CLS token layer $l$. 
With this combination type, average improvement over the baseline is achieved with the majority of early layers.
Specifically, we obtain accuracy improvements ranging from +0.2 to +2.0 for Devign in 8 out of 11 layers, 
and accuracy improvements from +0.1 to +0.8 for ReVeal in 9 out of 11 layers. 
The dynamic of the metric change over selected layer numbers is different for Devign and ReVeal.
In detail, the average performance of combination~(ii) is best with layer 3 on Devign (a +2.0 accuracy improvement) and with layer 1 for ReVeal (a +0.8 accuracy improvement). 
The best improvement in terms of F1(w) matches with layer 3 for Devign and with layer 2 for ReVeal, as shown in Figure~\ref{fig:one-layer-heatmap-f1}.

Max pooling over all available tokens from a selected layer in combination~(v) also achieves performance improvement over the baseline, 
as shown in rows 2 and 3 of Figures~\ref{fig:devign-one-layer-heatmap},~\ref{fig:reveal-one-layer-heatmap}. 
In general, layers 4--11 yield higher accuracy and layers 2--11 higher F1(w) with max pooling for Devign than the baseline. 
For ReVeal, all layers except layer 11 result in better average accuracy and layers 2--10 have higher average F1(w). 
Max pooling over all tokens, including special tokens, achieves the best statistically significant average improvement of accuracy of +0.9 of all combinations for ReVeal. 

The weighted sum of all tokens or code tokens exclusively in combination~(ix) does not improve the baseline performance. 
We assume that fine-tuning for 10 epochs is not enough for this type of combination, because the loss at epoch 10 on both training and validation splits is higher for combinations~(ix) than for combinations with max pooling. 
Since the goal of this study is to use the same or less resources for fine-tuning, we have not fine-tuned this combination for more than 10 epochs. %

While combinations~(ii) and (v) perform better for the majority of layers on the defect detection task, multi-class classification for bug or exception type prediction does not benefit from the combinations to the same extent as the binary task. 
Only max pooling of tokens of the last encoder layer achieves better performance than the baseline for BIFI (+0.1 accuracy, +0.1 weighted F1-score improvements) and Exception Type (a +0.2 accuracy, +0.1 weighted F1-score improvements) datasets. 

The impact of using all tokens or code tokens exclusively depends on the dataset. 
The difference between performance of single-layer combinations with max pooling of all tokens and only code tokens constitute 0.0-0.1 accuracy or F1(w).
For the multi-class tasks, the average results improve with the use of each later layer in the model.
We obtain performance improvement with the max pooling combination~(v), while other one-layer combinations do not perform better than the baseline. 

The best performing results on Devign and Exception Type classification datasets are statistically significant according to the Wilcoxon test. 
For ReVeal, the second best result is statistically significant.
We have not obtained statistically significant improvements for BIFI. 
We explain it by the fact that the baseline metric is already high, i.e., 96.7 accuracy.
Achieving improvement is usually more challenging when the baseline performs at this level.  

In essence, the combinations that involve \CLS tokens corresponding to the single layer~(ii), as well as the max pooling combinations~(v) perform better on average for defect detection datasets Devign and Reveal. 
However, only the max pooling combination~(v) of tokens from the last encoder layer outperforms the baseline on average for multi-class datasets BIFI and Exception Type. 
The weighted sum of tokens from a selected layer (ix) performs worse than the baseline if fine-tuned for the same number of epochs for all tasks. 
Multi-class classification tasks require the information from the last layer for better performance in our experiments, while the binary task of defect detection allows us to use early layers and improve the performance over the baseline.

\subsubsection{Multi-Layer Combinations}
\label{sec:all-layers}
\begin{figure}[t]
  \centering
  \includegraphics[width=\columnwidth, trim={0.3cm 0.3cm 0.28cm 0.2cm}, clip]{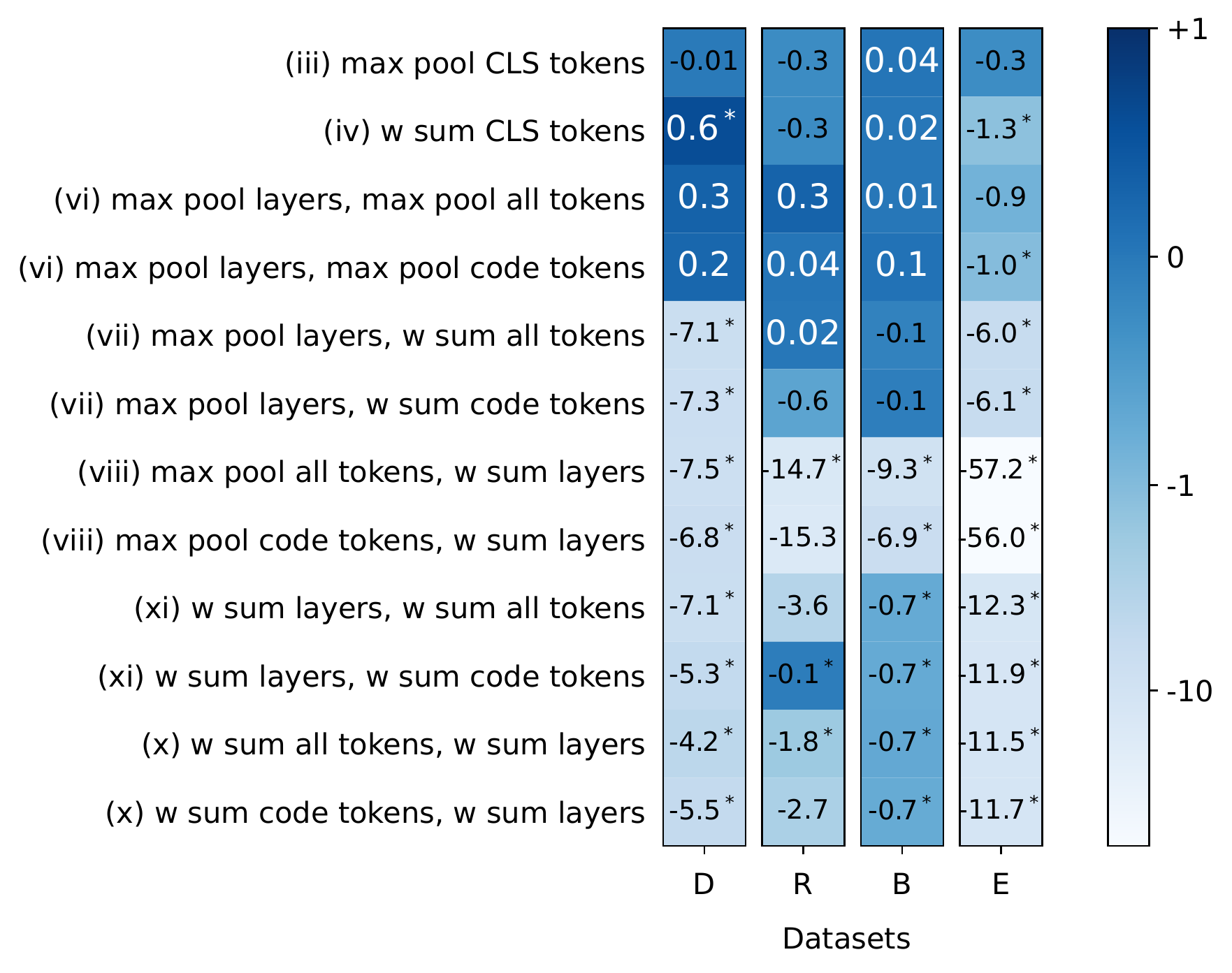}
  \caption{Difference of average accuracy between \method{} and baseline performance on D~(Devign), R~(ReVeal), B~(BIFI), E~(Exception Type).
  The star~$^*$ indicates a statistically significant difference w.r.t. the baseline.
  } 
\label{fig:all-token-heatmap}
\end{figure}

\begin{figure}[t]
\centering
\includegraphics[width=\columnwidth, trim={0.3cm 0.3cm 0.28cm 0.2cm}, clip]{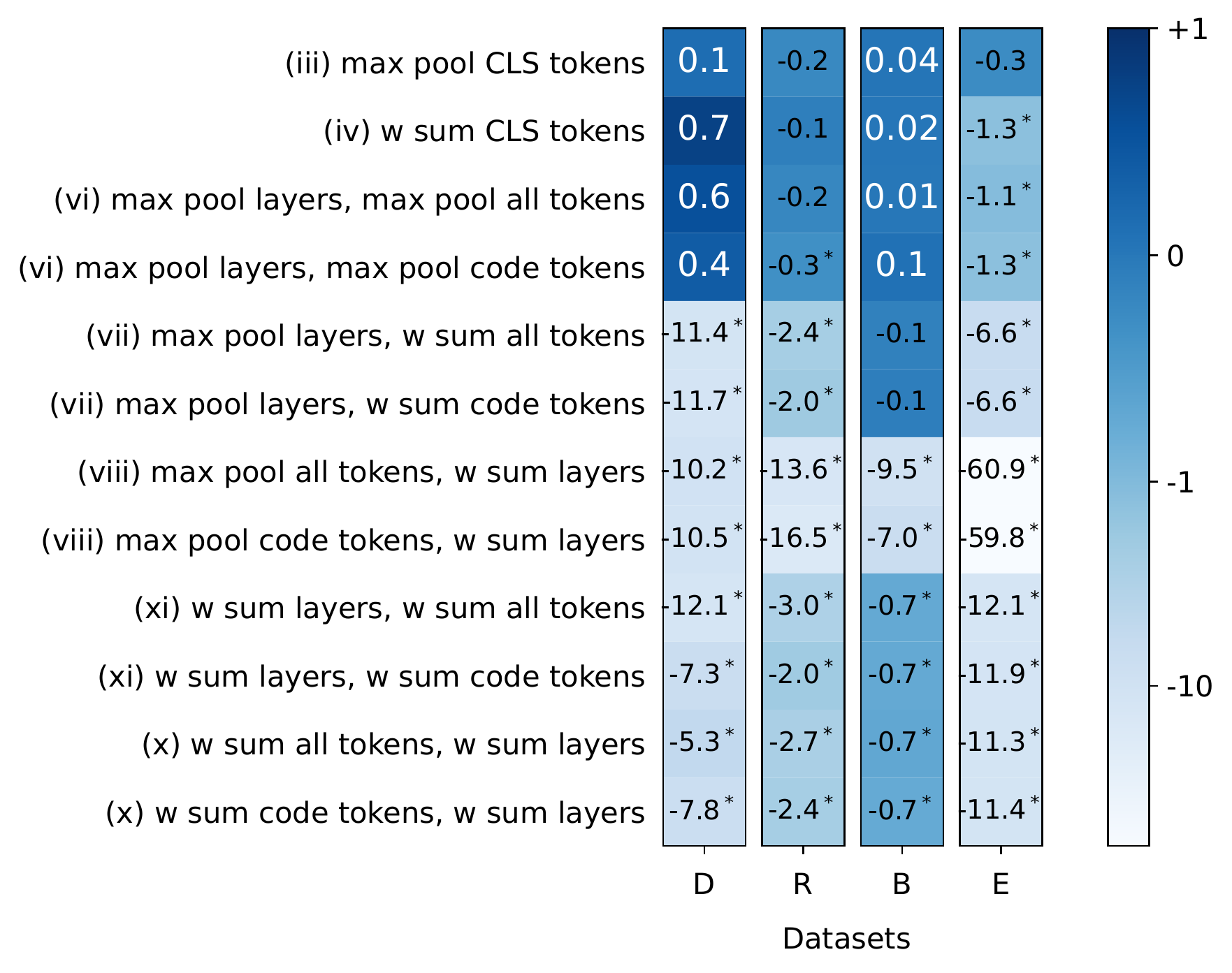}
\caption{Difference of mean weighted F1-score between combinations and baseline. 
Datasets are abbreviated to D~(Devign), R~(ReVeal), B~(BIFI), E~(Exception Type). 
The star~$^*$ indicates a statistically significant difference w.r.t. the baseline.
}
\label{fig:all-token-heatmap-f1}
\end{figure}

The average performance difference with the baseline of combinations that utilize 
early layers is shown \leon{as heatmaps} in Figures~\ref{fig:all-token-heatmap} and~\ref{fig:all-token-heatmap-f1}. 
We \leon{include the value of the average performance difference}
and add a star ($^*$) to the number if the 
\ana{difference}
is statistically significant. 
\leon{Again, negative values are shown in black, and positive values are shown in white.}

When we use all information from the available layers, the improvement over the baseline is less than what was observed in Section~\ref{sec:one-layer}, where one specific layer has been used. 
In detail, out of combinations that involve \CLS tokens from all early layers, no combination performs better than the baseline for ReVeal, BIFI, or Exception Type datasets.
However, the best improvement (+0.6 accuracy) out of experiments with all layers is obtained on Devign with the weighted sum of \CLS tokens in the combination~(iv), which is less than the maximum improvement with the combinations from one selected early layer in Section~\ref{sec:one-layer}. 
The improvement of F1(w) is shown in Figure~\ref{fig:all-token-heatmap-f1}. 
We obtained slightly better improvements of F1(w) for Devign, no F1(w) improvement for the unbalanced ReVeal dataset. 
The average F1(w) difference with the baseline for multi-class tasks are the same as accuracy difference.

If we consider the combinations that involve all tokens, the combination~(vi) with two max pooling operations outperforms the baseline for Devign, Reveal, and BIFI with accuracy improvement between +0.1 and +0.3. 
No combination that involves all layers outperforms the baseline on average for Exception Type dataset.
Combinations that involve one max pooling and one weighted sum of all tokens perform worse or neutral in comparison with the baseline. 
The combinations with only weighted sums perform worse than the baseline on average.

\begin{mdframed}[style=mystyle]
\noindent
\textbf{Answer to RQ1.} 
\method{} achieves statistically significant accuracy and F1-score improvements for defect detection datasets by using single-layer combinations that involve the \CLS token or max pooling over all tokens.
For bug type and exception type classification, max pooling of the tokens from the last encoder layer has improved the performance.
Weighted sum of tokens does not improve performance over the baseline. %
\end{mdframed}

\subsection{Pruned Models}

\begin{figure}[b]
\centering
\includegraphics[width=\columnwidth, trim={0.3cm 0cm 0.25cm 0}, clip]{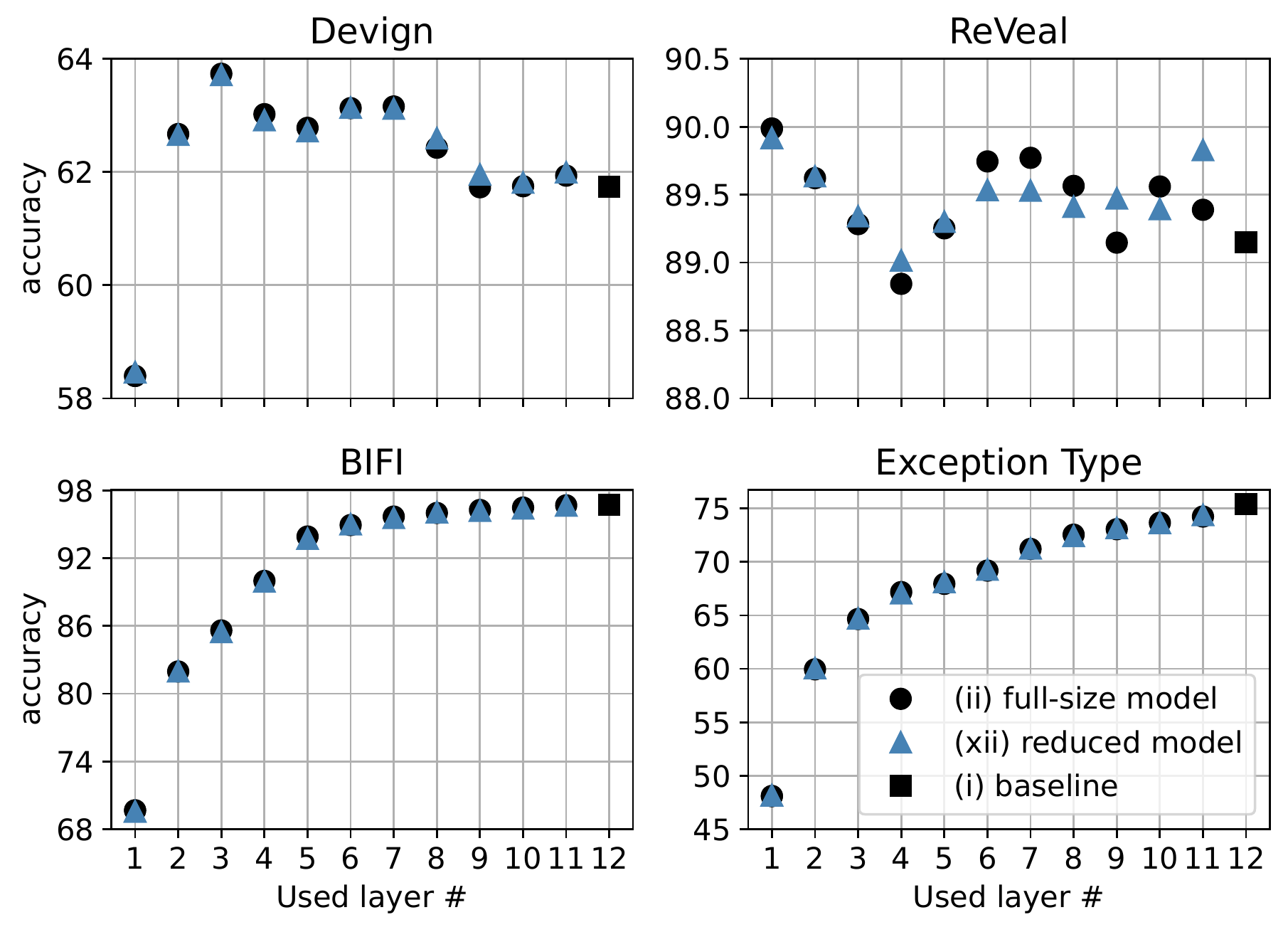}
\caption{Model performance with a subset of $l<L$ layers~(xii) vs. models with all layers~(ii); CLS token from layer $l$.}
\label{fig:one-layer-cls-diff-sizes}
\end{figure}

This section is devoted to the combinations of early layers that are initialized with the first $l<L$ early layers from the pre-trained model and fine-tuned as $l$-layer models --- combinations (xii). 
We start by comparing the performance of using the \CLS token from layer $l$ of the full-size model, i.e., combination~(ii), and using the \CLS token from layer $l$ of the model that has $l$ layers in total --- combination (xii). 
Figure~\ref{fig:one-layer-cls-diff-sizes} presents average accuracy obtained with these two combinations depending on the used layer, as well as the baseline combination of using  \CLS from the last layer $L=12$ of CodeBERT.  
On average, the pruned models with reduced size perform on par with the full-size model for defect detection on the balanced Devign dataset, and for bug type and exception type classification. 
However, the performance of the two analogical combinations diverges for the unbalanced defect detection dataset ReVeal in layers 4 and 6--11. 

\begin{table*}[t]
\newcommand{\tcbox}[2]{\colorbox{#1}{\raisebox{0pt}[1.1ex][0pt]{\hspace*{-2pt}#2\hspace*{-2pt}}}}
\caption{Comparison of reduced size models with the baseline. 
We report metric performance for the baseline and difference with the baseline for reduced models, 
average time for one-epoch fine-tuning (Time) in mm:ss format, 
speed-up and performance variation obtained with models with $l$ layers. 
Statistically significant improvements are marked in bold, 
statistically insignificant performance losses are marked with *, 
and ${A}_{12}$ effect sizes \ana{for accuracy and F1(w), if any,} are indicated by cell color, respectively \tcbox{RoyalBlue!50}{large}, \tcbox{RoyalBlue!25}{medium}, and \tcbox{RoyalBlue!10}{small}. 
The best metric improvement with highest speed-up factors are underlined.}
\label{tab:smaller-models}
\newcommand{\lef}{\cellcolor{RoyalBlue!50}}
\newcommand{\mef}{\cellcolor{RoyalBlue!25}}
\newcommand{\sef}{\cellcolor{RoyalBlue!10}}
\newcommand{\nef}{}
\centering
\begin{adjustbox}{max width=\textwidth}
\begin{tabular}{r|rcrr|rcrr|rcrr|rcrr}
\toprule
 & \multicolumn{4}{c|}{\bf{Devign}} & \multicolumn{4}{c|}{\bf{ReVeal}} & \multicolumn{4}{c|}{\bf{BIFI}} & \multicolumn{4}{c}{\bf{Exception Type}} \\
\multicolumn{1}{c|}{$l$} & Time & Speed-up & Acc & F1(w) & Time & Speed-up & Acc & F1(w) & Time & Speed-up & Acc & F1(w) & Time & Speed-up & Acc & F1(w)\\

\midrule
12     &                   8:50 &               1.0x &             61.7 &             60.4 &                   6:57 &               1.0x &             89.2 &             88.5 &                   8:41 &               1.0x &         96.7 &         96.7 &                   7:22 &              1.0x &         75.4 &         75.3 \\
\midrule
11.0     &                   8:03 &               1.1x &       \sef{+0.3} &            -0.1* &                   6:56 &               1.0x &  \lef{\bf{+0.6}} &       \mef{+0.3} &                   7:07 &               1.2x &  \sef{-0.1*} &  \sef{-0.1*} &                   6:33 &               1.1x &   \lef{-1.0} &   \lef{-1.0} \\
10.0     &                   7:13 &               1.2x &       \sef{+0.1} &       \sef{+0.3} &                   6:20 &               1.1x &       \sef{+0.2} &            -0.0* &                   6:22 &               1.4x &   \lef{-0.3} &   \lef{-0.3} &                   6:01 &               1.2x &   \lef{-1.8} &   \lef{-1.8} \\
9.0      &                   6:40 &               1.3x &       \sef{+0.3} &       \sef{+0.5} &                   5:53 &               1.2x &       \sef{+0.3} &      \sef{-0.1*} &                   5:46 &               1.5x &   \lef{-0.5} &   \lef{-0.5} &                   5:29 &               1.3x &   \lef{-2.2} &   \lef{-2.3} \\
8.0      &                   5:52 &               1.5x &  \lef{\bf{+0.9}} &       \lef{+1.1} &                   5:15 &               1.3x &       \sef{+0.2} &      \mef{-0.1*} &                   5:13 &               1.7x &   \lef{-0.6} &   \lef{-0.6} &                   4:55 &               1.5x &   \lef{-3.0} &   \lef{-3.0} \\
7.0      &                   5:23 &               1.6x &  \lef{\bf{+1.4}} &  \lef{\bf{+2.2}} &                   4:44 &               1.5x &             +0.3 &       \sef{+0.2} &                   4:43 &               1.8x &   \lef{-1.1} &   \lef{-1.1} &                   4:20 &               1.7x &   \lef{-4.1} &   \lef{-4.4} \\
6.0      &                   4:54 &               1.8x &  \lef{\bf{+1.4}} &  \lef{\bf{+2.0}} &                   4:25 &               1.6x &       \sef{+0.3} &             +0.1 &                   4:10 &               2.1x &   \lef{-1.7} &   \lef{-1.7} &                   3:50 &               1.9x &   \lef{-6.1} &   \lef{-6.6} \\
5.0      &                   4:03 &               2.2x &  \lef{\bf{+1.0}} &  \lef{\bf{+1.5}} &                   3:45 &               1.9x &       \sef{+0.1} &       \sef{+0.2} &                   3:31 &               2.5x &   \lef{-3.0} &   \lef{-3.0} &                   3:15 &               2.3x &   \lef{-7.3} &   \lef{-7.7} \\
4.0      &                   3:22 &               2.6x &  \lef{\bf{+1.2}} &  \lef{\bf{+1.6}} &                   3:06 &               2.2x &      \sef{-0.2*} &       \sef{+0.2} &                   2:53 &               3.0x &   \lef{-6.8} &   \lef{-6.8} &                   2:41 &               2.7x &   \lef{-8.3} &   \lef{-9.2} \\
3.0      &                   2:41 &               \underline{3.3x} &  \lef{\underline{\bf{+2.0}}} &  \lef{\underline{\bf{+2.4}}} &                   2:28 &               2.8x &             +0.1 &       \sef{+0.3} &                   2:15 &               3.8x &  \lef{-11.3} &  \lef{-11.3} &                   2:10 &               3.4x &  \lef{-10.7} &  \lef{-12.0} \\
2.0      &                   2:00 &               4.4x &  \lef{\bf{+1.0}} &  \lef{\bf{+1.9}} &                   1:52 &               \underline{3.7x} &       \lef{\underline{+0.4}} &  \lef{\underline{\bf{+0.6}}} &                   1:34 &               5.5x &  \lef{-14.7} &  \lef{-14.8} &                   1:34 &               4.7x &  \lef{-15.4} &  \lef{-17.1} \\
1.0      &                   1:18 &               6.8x &       \lef{-3.2} &       \lef{-2.3} &                   1:13 &               5.7x &  \lef{\bf{+0.8}} &       \lef{-1.2} &                   0:58 &               9.0x &  \lef{-27.2} &  \lef{-27.3} &                   0:59 &               7.4x &  \lef{-27.3} &  \lef{-30.7} \\
\bottomrule
\end{tabular}
\end{adjustbox}
\end{table*}

Most importantly, the results show that reducing the model size and using the \CLS token from the last layer of the reduced model 
performs on par with the baseline for the defect detection task. 
The best improvement with the reduced model is achieved with the 3-layer encoder for Devign and the 1-layer encoder for ReVeal.
This result shows that it is possible to both reduce resources and improve the model's performance during fine-tuning on the defect detection task with both a balanced and unbalanced dataset.

To explore the trade-off between resource usage and performance degradation for bug type and exception type identification, we show the average speed-up of one fine-tuning epoch and the performance loss compared to the baseline for BIFI and Exception Type datasets in Table~\ref{tab:smaller-models}.
We also report the corresponding values for Devign and ReVeal, for which both gains and losses of performance are indicated.
The speed up is reported as a scaling factor of the baseline time. 
The metric difference is shown as gain or loss of the weighted F1-score and accuracy compared to the baseline performance. 
Statistically significant improvements are reported in bold, while statistically insignificant losses are marked with a star~($^*$). 
\leon{The ${A}_{12}$ effect sizes are indicated by three shades of blue as the cell color, 
with the darkest shade indicating a large effect (${A}_{12} > 0.71$), 
the middle shade indicating a medium effect (${A}_{12} > 0.64$), 
and the lightest shade indicating a small effect (${A}_{12} > 0.56$).}
We also underline and discuss selected results that improve the metric values and reduce resource usage.

The majority of combinations~(xii) with pruned models outperform the baseline for Devign and ReVeal. 
Furthermore, models with 2--10 layers show statistically significant improvements of both metrics on Devign, with the 3-layer model achieving +2 accuracy improvement with a 3.3-times average speed-up of fine-tuning with the same hardware and software. 
Not only does the 3-layer model improve the accuracy over CodeBERT baseline to 63.7, but also outperforms several other models tested on Devign and reported on the CodeXGLUE benchmark~\cite{lu2021:codexglue}.
In particular, \leon{our pruned} 3-layer CodeBERT model outperforms the full-transformer model PLBART~\cite{ahmad2021:unified}, and code2vec code representations pre-trained on abstract syntax trees and code tokens in a joint manner~\cite{ahmad2021:unified}. \ana{However, our pruned model does not outperform the best performing model reported on CodeXGLUE, CoText, which achieves 66.62 accuracy~\cite{phan2021:cotext}.}

Models with 1 and 11 layers achieve statistically significant accuracy improvements for ReVeal. 
However, the 1-layer model reduces the F1(w) score. 
The use of layer 11 does not impact the speed of fine-tuning, while the 1-layer model yields the 3.7x acceleration of the baseline fine-tuning speed. 
The lack of speed-up with 11-layer model can be explained by the fact that the number of trainable parameters does not decrease linearly with the removal of later layers, since the additional embedding layer and classification head remain unchanged. 
The 2-layer model results in the best improvement of F1(w) which is statistically significant. The 2-layer model improves accuracy on ReVeal as well. 
\ana{For Devign and ReVeal, statistically significant improvements have large effect size.}

For BIFI, we obtain statistically insignificant decrease of F1(w) and accuracy according to the Wilcoxon test which brings about 1.2x speed-up of the fine-tuning with the 11-layer model. 
If we decrease the number of layers to 8, the performance on BIFI stays within the (baseline metric$-1$) limit, but we gain up to 1.7x average speed-up of one-epoch fine-tuning.
In case of using models with 1--10 layers, we observe a statistically significant change of distribution and decrease of metric values.

For the unbalanced Exception Type dataset, the performance drops faster and the speed-up is less prominent than for BIFI. 
The change of mean values of the metrics for all models is statistically significant. 
In detail, the metrics decrease by -1.0 absolute metric value at 11 layers with 1.1x fine-tuning speed-up and by -1.8 with 10 layers with 1.2x speed-up.
We explain the sharper decline of the combinations performance by the lower baseline metric values (75.39 accuracy, 75.30 weighted F1-score) than in the case of BIFI (96.7 accuracy and weighted F1-score). 
\ana{For BIFI, statistically insignificant deterioration have small effect size. 
However, for both BIFI and Exception Type datasets, we observe deterioration of performance of large effect size with pruned models.}

We conclude that for the BIFI dataset with high-performing baseline and 3 classes, the performance loss at removing each layer is less than for the Exception Type classification dataset with lower baseline performance and 20 classes. 
The resource usage, which is correlated with time spent on tuning, decreases faster for BIFI than for Exception Type. 
This is partially explained by a larger classification head for the Exception Type dataset, because this dataset has 20 classes as opposed to only 3 classes in BIFI. 
\ana{In other words, we observe that the BIFI dataset has a strong baseline that is hard to outperform with pruning.
By contrast, the complexity of the Exception Type dataset can influence the results in the opposite way: \leon{The baseline performance is already not very strong, and it proves hard to further improve on it with early layers only.}}

\begin{mdframed}[style=mystyle]
\noindent
\textbf{Answer to RQ2.} 
We obtain performance improvements over the baseline as well as fine-tuning speed-ups for both defect detection datasets by using the {\CLS} token from the last layer of pruned models. 
For multi-class classification, performance decreases upon pruning each layer from the end of the model. 
The decrease is sharper for the dataset with 20 exception types than for the task with 3 bug types.
\end{mdframed}

\subsection{Threats to Validity}
\noindent
The main threat to external validity is that the results are empirical and may not generalize to all code classification settings, 
including other programming languages, tasks, and encoder-based models for code. 
We have tested \method{} combinations on code in C for defect detection and Python for bug type and exception type classification in this study. 
The choice of the CodeBERT as the encoder model and its internal structure affects the results. 
For instance, an encoder model that takes smaller input sequences can perform worse on the same datasets, because larger parts of input code sequences have to be pruned in this case.
The external validity can be improved by testing on more datasets and encoder models.

The threats to internal validity concern the dependency of models on initializations of trainable parameters and the choice of methods. 
Classification head and weighted sums with trainable parameters in our experiments depend on the initialization of the parameters and can lead the model to arrive at different local minima during fine-tuning.  
To reduce the effect of different random initializations, we have fine-tuned and tested all \method{} combinations 10 times with different random seeds. 

In addition, we used the Wilcoxon test to verify whether the achieved improvements are statistically significant. 
However, the Wilcoxon test only estimates whether measurements of baseline values and \method{} combinations are drawn from different distributions. 
The reported times spent on fine-tuning and corresponding speed-ups have the purpose of illustrating the reduction in resource usage, and will depend on the hardware used. %
Even when using factors of speed-up for pruned models, there is a chance that these numbers will be different on other hardware configurations.  

We implemented the algorithms and statistical procedures in Python, with the help of widely used libraries such as PyTorch, NumPy and SciPy. 
However, we cannot guarantee the absence of implementation errors which may have affected our evaluation.

\section{Concluding Remarks}
\label{sec:conclusion}

In this paper, we have proposed \method{}, 
an approach to combine early layers of encoder models for code, 
and tested different early-layer combinations on the software engineering tasks of defect detection, 
bug type and exception type classification. 
Our study is motivated by the hypothesis that early layers contain valuable information that is discarded by the standard practice of representing the code with the \CLS token from the last encoder layer. 
\method{} provides ways to improve the performance of existing models with the same resource utilization, 
as well as for resource usage reduction while obtaining comparable results to the baseline.  

\head{Results}
Using \method{}, we obtain statistically significant improvements over the baseline %
for the majority of the combinations that involve a single encoder layer on defect detection, 
and with selected \method{} combinations on bug type and exception type classification. 
Max pooling of tokens from selected single layers yields performance improvements for all datasets.
Both the classification performance and the average fine-tuning time for one epoch are improved by 
pruning the pre-trained model to its early layers and using the \CLS token from the last layer of the pruned model.
For defect detection, this results in a +2.0 increase in accuracy and a 3.3x fine-tuning speed-up on Devign, and up to +0.8 accuracy improvement with a 3.7x speed-up on ReVeal. 
Pruned models do not lead to multi-class classification performance gains, but they do show a fine-tuning speed-up and the associated decrease in resource consumption. 

The results show that pruned models with reduced size either work better or can result in a reduction of resource usage during fine-tuning with different levels of performance variation, 
which indicates the potential of \method{} in resource-restricted scenarios of deploying defect detection and but type classification in production environments. 
For example, \method{} achieves a 2.1x speed-up for BIFI while reducing accuracy from 96.7 to 95.0.

\head{Future Work}
The study can be extended by investigating the generalization to other encoder models.
\ana{We are in the process of studying the performance of \method{} with two new encoder models: StarEncoder~\cite{li2023:starcoder} and ContraBERT\_C~\cite{liu2023:contrabert}.}
\leon{Another direction for future research is whether the types of layer combination and pruning, as we have investigated in this paper for encoder architectures, are also effective techniques for decoder and encoder-decoder architectures.}
Moreover, it would be of interest to experiment with other code classification tasks, such as general bug detection and the prediction of vulnerability types. 
The latter could be investigated using the CWE types from the Common Weakness Enumeration as labeled in the CVEfixes dataset~\cite{bhandari2021:cvefixes}. 

\begin{acks}
This work is supported by the Research Council of Norway through the secureIT project (IKTPLUSS \#288787).
Max Hort is supported through the ERCIM ‘Alain Bensoussan’ Fellowship Programme.
The empirical evaluation presented in this paper was performed on the Experimental Infrastructure for Exploration of Exascale Computing (eX3), 
financially supported by the Research Council of Norway under contract \#270053, 
as well as on resources provided by Sigma2, 
the National Infrastructure for High Performance Computing and Data Storage in Norway.
\end{acks}

\section*{Data Availability}
To support open science and allow for reproduction and verification of our work, all artifacts are made available through Zenodo at the following URL: \url{https://doi.org/10.5281/zenodo.7608802}. 

\bibliographystyle{ACM-Reference-Format}
\bibliography{arXiv_Grishina_EarlyBIRD} 

\end{document}